\documentclass[a4paper,12pt]{article}
\usepackage{jheppub} 
\usepackage{lineno}

\usepackage[english,activeacute]{babel}
\usepackage{mathtools}
\usepackage[scr=boondox,  
            cal=esstix]   
           {mathalpha}

\title{\boldmath Weyl double copy in bimetric massive gravity}

 \author[1]{Hugo Garc\'{\i}a-Compe\'an\note{hugo.compean@cinvestav.mx}}
 \author[2]{and C\'esar I. Ramos\note{cesar.ramos@cinvestav.mx}}
 \affiliation{Departamento de F\'{\i}sica, \\ 
 Centro de Investigaci\'on y de Estudios Avanzados del Instituto Polit\'ecnico Nacional,\\
 P.O. box 14-740, C.P. 07000, Ciudad de M\'exico, Mexico}

\abstract{The Weyl double copy formalism, which relates the Weyl spinor with the square of the field strength, is studied in the context of Hassan-Rosen bigravity for stationary and time-dependent solutions. We consider the dyonic Kerr-Newman-(A)dS solution and the Pleba\'nski-Demia\'nski metric in the context of bigravity. These solutions are studied in the Weyl double copy both with matter independently coupled and show that no massive modes are present in the Weyl spinor. The equations of motion for the gauge and scalar fields are those of Maxwell equations coupled to an external source, and massless Klein-Gordon equations with a conformal curvature term and an external source, all of them consistent with general relativity. For wave solutions, massive modes are manifest in the Weyl spinor and a formulation in bigravity for these massive modes is proposed. The resulting equations of motion are Proca equations with a conformal term and massive Klein-Gordon equations. In the case of the matter contributions for waves, we show how the resonance mass is present in equations of motion of the fields obtained from the Weyl double copy. The solutions studied are written in a Kerr-Schild form, connecting with the Kerr-Schild double copy.}

\makeatletter
\gdef\@fpheader{}
\makeatother

\begin{document}
\maketitle
\flushbottom
\section{Introduction}

There are several interesting proposals connecting gravity and gauge theories. Among them, the string duality constructions, the Gauge/Gravity holographic duality and the Double Copy conjecture seems to be the most studied in the literature \cite{Polchinski:2014mva,Maldacena:2011ut,Bern:2010ue}. In the Double Copy proposal one considers a
relation between color and kinematic factors in the scattering amplitudes of gauge and gravity theories \cite{Bern:2008qj,Bern:2010yg,Bern:2010ue}. This relation is rooted in the link between amplitudes of closed and open strings in string theory \cite{Kawai:1985xq}. This phenomenon also has a classical counterpart, which relates classical solutions of gravity theories to solutions of gauge theories. The Kerr-Schild double copy maps the Kerr-Schild solutions in general relativity (GR) to solutions in gauge theories \cite{Monteiro:2014cda,Luna:2015paa,Luna:2016due,Bahjat-Abbas:2017htu,Carrillo-Gonzalez:2017iyj,Alkac:2021bav,Kerr:1965wfc,Gurses:1975vu,Stephani:2003tm}. For example, in \cite{Monteiro:2014cda}, the Schwarzschild black hole is related to the Coulomb point-like solution via this procedure. Another classical double copy relation is the Weyl double copy (WDC) \cite{Luna:2018dpt,Godazgar:2020zbv,Easson:2021asd,Easson:2022zoh}, which we will study in this work. 

The Weyl double copy (WDC) \cite{Luna:2018dpt} is a relation between classical gravity and gauge theory solutions where the Weyl tensor is written using the square of a field strength tensor. It considers algebraically special solutions and the spinorial formalism of GR. Here, gravity solutions are mapped to Maxwell solutions, which can be seen as the abelian version of Yang-Mills, and they are related to the result in the amplitudes double copy. For general vacuum Petrov type {\bf D} solutions, the Weyl double copy was studied in \cite{Luna:2018dpt} and for type {\bf N} solutions in \cite{Godazgar:2020zbv}. A generalization that considers sources was given in \cite{Easson:2021asd} and a complete analysis of electrovacuum type {\bf D} solutions was given in \cite{Easson:2022zoh}, which we will be referring to as the sourced Weyl double copy (SWDC). This gives a framework for relating gravity theory solutions to gauge and scalar theory solutions.

On the other hand, driven by the observations (e.g., dark matter, dark energy) than GR has not been able to fully explain, in the literature there are been various proposals of modifying GR in order to explain such phenomena. From the particle physics perspective, GR is the  unique theory of a spin-2 massless particle, the graviton, which couples to matter, is Lorentz invariant and preserves locality \cite{Armas:2021yut,Gupta:1954zz,Weinberg:1965rz,Weinberg:1980kq, Wald:1986bj,Feynman:1996kb}. One option is to modify the zero-mass assumption of the graviton, thus modifying the behavior of the gravitational interaction. dGRT massive gravity \cite{deRham:2010ik,deRham:2010kj} is a  ghost-free theory with these characteristics, which introduces the massive term by means of an auxiliary metric. The proposal which gives dynamics to the fiducial metric is bigravity \cite{Hassan:2011zd} and for multiple metrics we have multi-metric massive gravity \cite{Hinterbichler:2012cn,Hassan:2018mcw,Flinckman:2024zpb}. WDC relations can be studied for generalizations of GR, which can answer to which gauge theories these alternatives of GR are connected, at least at a classical level. This could serve as a guiding principle when considering the duality at the amplitude level.

In the present work we study Hassan-Rosen bigravity \cite{Hassan:2011zd}, which considers two interacting spin-2 fields, one massive and one massless, and which endows the theory with features not present in GR, interesting from the phenomenological point of view \cite{Babichev:2016bxi,Babichev:2016hir,GonzalezAlbornoz:2017gbh,Marzola:2017lbt, deRham:2014fha, Hassan:2014gta,Comelli:2015pua,Hogas:2021fmr,Hogas:2021lns,Hogas:2021saw}. We aim to study the gauge and scalar theory related to this theory of massive gravity by pointing towards the Weyl double copy framework.

The article is outlined as follows: in Sec. \ref{Sec_WDC in GR} we consider the spinorial formalism of GR and the Weyl double copy relations, specifically for type {\bf D} and type {\bf N} solutions, and then show some examples. In Sec. \ref{Sec_WDC in BG}, we briefly overview Hassan-Rosen bigravity and the matter coupling in the theory, followed by an analysis of the Weyl spinor in the Weyl double copy in bigravity. After that we present the equations of motion for type {\bf D} and type {\bf N} solutions in bigravity, finishing the section with examples of solutions in the theory. We finish with the conclusions and perspectives in Sec. \ref{Sec_conclusions}.

\section{Weyl double copy in general relativity}\label{Sec_WDC in GR}
\subsection{Spinor formalism}

General relativity permits a formulation in terms of spinors \cite{Penrose:1960eq}. The objects used to transform tensors to spinors are the Infeld-van der Waerden symbols $\sigma_{A\dot{A}}^\mu$, defined using the Pauli matrices $\sigma_{i}$ with  $i=\{x,y,z\}$. Then, by choosing a tetrad $e^{a}{}_\mu$, the ``spinorial vierbein'' $\sigma_{A\dot{A}}^\mu$ is defined as:
\begin{align}
    \sigma_{A\dot{A}}^\mu=(e^{-1})_a^\mu\,\sigma_{A\dot{A}}^a\,,\qquad \widetilde{\sigma}^{\mu\dot{A}A}=(e^{-1})_{a}^{\mu}\widetilde{\sigma}^{a\dot{A}A}\,,
\end{align}
where we have also defined $\widetilde{\sigma}^{\mu\dot{A}A}$ for later use. Tensor indices are represented with greek letters $\mu,\nu,\ldots$ and are raised/lowered using the metric, $g_{\mu\nu}$, whereas for tangent space indices we use lowercase latin letters $a,b,\ldots$  and use the flat tangent space metric $\eta_{ab}$ to lower and raise indices. Spinor indices are labeled with uppercase latin letters $A,\dot{A},B,\dot{B},\ldots$.  For spinor indices we employ uppercase latin letters $A,\dot{A},B,\dot{B},\ldots$ and we manipulate them by means of the ``epsilon spinor'', $\varepsilon_{AB}$, which for our convention is:
\begin{align*}
\varepsilon_{AB}=
\begin{pmatrix}
0 & 1 \\
-1 & 0
\end{pmatrix}_{AB},\quad\varepsilon^{AB}=
\begin{pmatrix}
0 & 1 \\
-1 & 0
\end{pmatrix}^{AB}\, ,\quad 
\varepsilon_{\dot{A}\dot{B}}=
\begin{pmatrix}
0 & 1 \\
-1 & 0
\end{pmatrix}_{\dot{A}\dot{B}},\quad\varepsilon^{\dot{A}\dot{B}}=
\begin{pmatrix}
0 & 1 \\
-1 & 0
\end{pmatrix}^{\dot{A}\dot{B}}\,.
\end{align*}
We also define $\varepsilon_{\dot{A}\dot{B}}$ and $\varepsilon^{\dot{A}\dot{B}}$, used to raise and lower the anti-self-dual indices, e.g., $V^{A\dot{A}}=V_{B\dot{B}}\varepsilon^{AB}\varepsilon^{\dot{B}\dot{A}}, V_{A\dot{A}}=V^{B\dot{B}}\varepsilon_{BA}\varepsilon_{\dot{B}\dot{A}}$. The inverse metric is defined as $g^{\mu\nu}=(e^{-1})_a^\mu(e^{-1})_b^\nu\eta^{ab}$, while $\eta^{ab}=\eta_{ab}$.

Depending on the signature of the solution we consider, the definition of $\sigma^{a}_{A\dot{A}}$ will differ. We work in GR and bigravity with solutions in $3+1$ spacetime dimensions with a 'mostly minus' signature, so that the flat tangent space metric is $\eta_{ab}=\mathrm{diag}(1,-1,-1,-1)$. For dimension $2+2$, we use $\eta_{ab}=\mathrm{diag}(1,1,-1,-1)$. The corresponding $\sigma^{a}_{A\dot{A}}$ for both cases is expressed as follows:
\begin{align} 
    \nonumber &\sigma^a_{A\dot{A}}=-\frac{1}{\sqrt{2}}\left({\bf 1},\sigma_x,-\sigma_y,\sigma_z\right)_{A\dot{A}}\,,\quad  \widetilde{\sigma}^{a\dot{A} A}=\frac{1}{\sqrt{2}}\left({\bf 1},-\sigma_x,-\sigma_y,-\sigma_z\right)^{\dot{A}A}\,;\\
    \nonumber  \quad & \sigma^a_{A\dot{A}}=\frac{1}{\sqrt{2}}\left({\bf 1},-i\,\sigma_y,\sigma_x,\sigma_z\right)_{A\dot{A}}\,, \quad \widetilde{\sigma}^{a A\dot{A}}=\frac{1}{\sqrt{2}}\left(-{\bf 1},i\,\sigma_y,\sigma_x,\sigma_z\right)^{\dot{A}A}\,,
\end{align}
where ${\bf 1}$ is the identity $2\times 2$ matrix. The first line is for Lorentzian signature and the second convention corresponds to Kleinian signature. We see that these objects satisfy some identities, such as:
\begin{align}
    \nonumber \sigma_{\mu}^{A\dot{A}}=g_{\mu\nu}\sigma_{B\dot{B}}^{\nu}\varepsilon^{BA}\varepsilon^{\dot{B}\dot{A}},\quad\sigma_{A\dot{A}}^{\mu}\sigma_{\nu}^{A\dot{A}}=\delta_{\nu}^{\mu}\,,\quad\sigma_{A\dot{A}}^{\mu}\sigma_{\mu}^{B\dot{B}}=\delta_{A}^{B}\delta_{\dot{A}}^{\dot{B}}\,,\quad \sigma_{A\dot{A}}^{a}\sigma_{b}^{A\dot{A}}=\delta_{b}^{a}\,.
\end{align}
These tools permit us to convert a $n$-index tensor $V_{\alpha\beta\cdots\gamma}$ to a $2n$-index spinor $V_{A\dot{A}B\dot{B}\cdots C\dot{C}}$:
\begin{align*}
    V_{\alpha\beta\cdots\gamma} \mapsto V_{A\dot{A}B\dot{B}\cdots C\dot{C}}= V_{\alpha\beta\cdots\gamma}\,\sigma_{A\dot{A}}^{\alpha}\,\sigma_{B\dot{B}}^{\beta}\cdots\sigma_{C\dot{C}}^{\gamma} \,.
\end{align*}
Within this framework, the spinorial form of the Weyl tensor is
\begin{align}
    W_{A\dot{A}B\dot{B}\dot{C}\dot{C}D\dot{D}}=C_{ABCD}\varepsilon_{\dot{A}\dot{B}}\varepsilon_{\dot{C}\dot{D}}+\overline{C}_{\dot{A}\dot{B}\dot{C}\dot{D}}\varepsilon_{AB}\varepsilon_{CD}\,.
\end{align}
In the Weyl double copy, we work with the self-dual part of the spinor, which can be obtained by using the  two-index antisymmetric tensor $\sigma_{AB}^{\mu\nu}$, written as
\begin{align}
    C_{ABCD}=\frac{1}{4}W_{\alpha\beta\mu\nu}\sigma_{AB}^{\alpha\beta}\sigma_{CD}^{\mu\nu}\, , \qquad \sigma_{AB}^{\mu\nu} = \sigma_{A\dot{A}}^{[\mu}\widetilde{\sigma}^{\nu]\dot{A}C}\varepsilon_{CB}\,.
\end{align}
The field strength $F_{\mu\nu}$ can be mapped into a 4-index spinor tensor, $F_{A\dot{A}B\dot{B}}$, decomposed as
\begin{align}
\label{Fmunu_spinorial}
    F_{A\dot{A}B\dot{B}} = f_{AB}\varepsilon_{\dot{A}\dot{B}}+\overline{f}_{\dot{A}\dot{B}}\varepsilon_{AB}\, , \ \ \ \quad 
    f_{AB}=\frac{1}{2}F_{\mu\nu}\sigma_{AB}^{\mu\nu}.\, 
\end{align}
The definitions we have presented here will be useful to construct and operate the objects that involve spinors in the Weyl double copy prescription. 

Weyl double copy relations involve the use of Weyl invariants and principal null directions, which are directions where the Weyl tensor has a special behavior. We have to choose a tetrad in the formalism, and not every tetrad is useful in this context. We need to choose a tetrad aligned with the principal null directions of the Weyl tensor, and it is in such tetrad where the Weyl invariants are physically meaningful. We employ tetrads aligned to the principal null direction of the solution we are considering in each case.
 
\subsection{Weyl double copy}

The Weyl double copy, as presented in \cite{Luna:2018dpt} for type {\bf D} vacuum metrics, is given by
\begin{align}
    \label{WDC_1}
    \Psi_{ABCD}=\frac{1}{S_{(g_{1})}}f^{(g_1)}_{(AB}\cdot f^{(g_{1})}_{CD)} \,, \qquad \,^{(0)}\nabla_{\mu}F_{(g_{1})}^{\mu\nu}=0\, , \quad \,^{(0)}\Box S_{(g_1)}=0,\,
\end{align}
where $\Box \equiv g^{\mu\nu}\nabla_{\mu}\nabla_{\nu}$ and the superscript ``${(0)}$'' denotes that the object is constructed with the flat metric, obtained by taking a certain limit. $S_{(g_1)}$ is the zeroth copy and the spinorial field strength $f^{(g_1)}_{AB}$ is related to the gauge field, i.e., with the single copy field, and $F^{(g_1)}_{\mu\nu}$ is its tensorial form. The Weyl double copy for type {\bf N} spacetimes was studied in \cite{Godazgar:2020zbv}, where a decomposition and equations as in (\ref{WDC_1}) hold. We add the subscript $A_{(g_i)}$ because in bigravity these expressions will be generalized to two metrics $G=\{g,f\}$, i.e., $A_{(G_i)}=\{A_{(g_i)},A_{(f_i)}\}$.

Considering external sources in order to tackle, for example, charged black hole solutions, we need to extend this formalism to the one proposed in \cite{Easson:2021asd,Easson:2022zoh}, namely, the sourced Weyl double copy (SWDC). In SWDC, the Weyl spinor $\Psi_{ABCD}$ is constructed as a sum different contributions of gauge and scalar theories and the fields $f^{(g_{i})}_{AB}$ and $S_{(g_i)}$ satisfies
\begin{align}
    \nonumber \Psi_{ABCD}=\sum_{i=1}^N\frac{1}{S_{(g_i)}}f_{(AB}^{(g_i)} \cdot f_{CD)}^{(g_i)} \,, \qquad \,^{(0)}\nabla_{\mu}{F_{(g_i)}}^{\mu\nu}=\,^{(0)}J_{(g_i)}^{\nu}\, , \quad \,^{(0)}\Box S_{(g_i)}=\,^{(0)}j_{(g_i)},
\end{align}
where $\,^{(0)}J_{(g_i)}^\mu$ and $\,^{(0)}j{_{(g_i)}}$ are external sources for the single and zeroth copies respectively. For $i=1$, we have the vacuum part and the sources vanish, whereas the value that corresponds to the external source, $i=2$, has a non-vanishing source term. We will be referring to these terms as the massless and the matter contributions respectively.

We can further extend the framework to maximally symmetric backgrounds, i.e., flat and (A)dS spacetimes, which can turn on the cosmological constant $\Lambda$. Even though the applicability of the double copy in curved backgrounds in the amplitudes story is still under study, the classical double copy has been shown to hold in curved backgrounds \cite{Han:2022mze}, where type {\bf N} and type {\bf D} solutions are shown to satisfy conformally invariant field equations. Then, when we write the equations of motion, we will use a bar ``$\overline{A}$'' on top of the quantities and operators that are associated with constant curvature backgrounds, incorporating AdS and dS spacetimes. To set the stage for the work in massive gravity, we focus on type {\bf N} and type {\bf D} solutions in a maximally symmetric background, analyzing the behavior of the Weyl spinor in both cases. 

The scalar and gauge theory will live in a maximally symmetric background, so that the field strength $f^{(g_i)}_{AB}$ will be obtained by using $\,^{(\overline{0})}\sigma_{AB}^{\mu\nu}$ in (\ref{Fmunu_spinorial}), which is constructed by taking the corresponding limit to the background metric $\overline{g}_{\mu\nu}$ instead of $\sigma_{AB}^{\mu\nu}$ of the full metric. The symbols $\,^{(\overline{0})}\sigma_{AB}^{\mu\nu}$ use the tetrad $\,^{(\overline{0})}{e_{g}}^{a}_{\mu}$ constructed by taking the same limit to the background metric in ${e_{g}}^{a}{}_{\mu}$. For a flat background, we use $\,^{(0)}{e_{g}}^{a}_{\mu}$ and $\,^{(0)}\sigma_{AB}^{\mu\nu}$.

\subsection{WDC of Type N and Type D solutions in GR}
Within the spinorial framework in GR, the ten independent components of the Weyl tensor can be written using five independent complex scalars. We focus on two of them for which the Weyl spinor can be written as
\begin{align}
    \label{typeN_D_GR}
    \nonumber &\text{Type {\bf N}:} \qquad \,^{(g)}\Psi_{ABCD}=\psi_{4}o_{A}o_{B}o_{C}o_{D}\, , \ \ \qquad \,^{(g)}\psi_{4}=\Psi_{ABCD}\iota^{A}\iota^{B}\iota^{C}\iota^{D}\,,\\
    &\text{Type {\bf D}:} \qquad \,^{(g)}\Psi_{ABCD}=6\psi_{2}o_{(A}o_{B}\iota_{C}\iota_{D)} \,, \ \ \quad 
    \,^{(g)}\psi_{2}=\Psi_{ABCD}o^{A}o^{B}\iota^{C}\iota^{D}\, .
\end{align}
 We again add the superscript ``$(g)$'' which will be later generalized for both metrics in bigravity. The spinor basis $\{o_{A},\iota_{B}\}$ we use is
\begin{align} 
    \nonumber  o_{A}=\frac{1}{\sqrt{2}}\left(1,1\right)_{A}\,,\, \, \iota_{A}=\frac{1}{\sqrt{2}}\left(1,-1\right)_{A}\, ; \quad \ 
     o_{A}=\frac{1}{\sqrt{2}}\left(1,-i\right)_{A}\,,\,\, \iota_{A}=\frac{1}{\sqrt{2}}\left(i,-1\right)_{A}\,,
\end{align}
such that $o^{A}\iota^{B}\varepsilon_{AB}=1$. The first basis is used when dealing with solutions with Lorentzian signature and the second corresponds to Kleinian signature solutions.

We construct the gauge and scalar theory solution by identifying
\begin{align}
    \label{def_gauge_scalar_th}
     \nonumber \text{Type {\bf N}}: & \ \ \ \  f^{(g_i)}_{AB}=\Theta_{g_i}\, o_{A}o_{B},\,  \quad S_{(g_i)}=\frac{(\Theta_{g_i})^{2}}{\psi_{g_i}}\,, \quad \,^{(g)}\psi_{4}=\sum_{i}\psi_{g_i}, \\
    \text{Type {\bf D}}:& \ \ \ \ f^{(g_i)}_{AB}=\Theta_{g_i} \,o_{(A}l_{B)}\, , \quad S_{(g_i)}=\frac{(\Theta_{{g_i}})^{2}}{6\psi_{g_i}}\,, \quad \,^{(g)}\psi_{2}=\sum_{i}\psi_{g_i}\,.
\end{align}
As before, $i=1$ corresponds to the vacuum contribution and $i=2$ to the matter contribution, as in \cite{Easson:2021asd}. In each case we decompose $\psi_{g_i}$ as the sum of vacuum and matter parts. For type {\bf N} solutions we use $\psi_4=\psi_{g_1}+\psi_{g_2}$ and for type {\bf D} solutions $\psi_2=\psi_{g_1}+\psi_{g_2}$.

For type {\bf D} solutions we work with the Kerr-Newman solution and the general electrovacuum solution of this type, the Pleba\'nski-Demia\'nski metric. For these solutions, the massless and matter fields obtained from the Weyl double copy obey
\begin{align}
\label{typeD_eom_1} 
    \nonumber \overline{\nabla}_{\mu}F_{(g_1)}^{\mu\nu}&=0\, , \quad &&\overline{\Box}\, S_{(g_1)}-\frac{\overline{R}}{6} S_{(g_1)}=0, \,\\
    \overline{\nabla}_{\mu}F_{(g_2)}^{\mu\nu}&=J^{\nu}_{(g_2)}\, , \quad
    &&\overline{\Box}\, S_{(g_2)}-\frac{\overline{R}}{6} S_{(g_2)}=j_{(g_2)}\,,
\end{align}
where 
\begin{align*}
    J^{\nu}_{(g_2)}=\,^{(0)}J_{(g_2)}^{\mu}+\widetilde{\Lambda}_{(g_2)}^\nu\,,\qquad \ \ \ j_{(g_2)}=\,^{(0)}j_{(g_2)}+\widetilde{\lambda}_{(g_2)}, 
\end{align*}
and $\,^{(0)}J_{(g_2)}^{\mu}$ and $\,^{(0)}j_{(g_2)}$ are related to the gravitational energy density of the background spacetime and depend on the matter terms, while $\widetilde{\Lambda}_{(g_2)}^\nu$ and $\lambda_{(g_2)}$ source the cosmological constant in each case and vanish for flat backgrounds. We interpret these terms as external sources in the gauge and scalar theories. These equations coincide with \cite{Han:2022mze,Easson:2021asd} in the flat background and vacuum limits respectively. 

For type {\bf N} vacuum solutions, in \cite{Godazgar:2020zbv}, the authors obtained that the massless fields of the Robinson-Trautman ($\Lambda$) solutions obey the same equations as in (\ref{typeD_eom_1}). In this paper we work with Siklos-(A)dS solutions, where the equations of motion for the fields in the Weyl double copy are found to be
\begin{align}
\label{typeN_eom_1} 
   \nonumber  &\overline{\nabla}_{\mu}F_{(g_1)}^{\mu\nu}- \frac{\overline{R}}{6} A^{\nu}_{(g_1)}=0\, , \qquad &&\overline{\Box}\, S_{(g_1)}=0\,,\\
    &\overline{\nabla}_{\mu}F_{(g_2)}^{\mu\nu}- \left(\frac{\overline{R}}{6}+m^{2}_{\text{res}}\right) A^{\nu}_{(g_{2})}=0\, , \qquad &&\left(\overline{\Box}+m^{2}_{\text{res}}\right)\, S_{(g_2)}=0\,,
\end{align}
where $m_\text{res}$ is the resonance mass,  i.e., the mass in which the external field (matter) inhomogeneity enters in resonance with the vacuum modes, and it depends on the matter content of the solution. pp-waves satisfy the vacuum equations for vanishing cosmological constant. Both equations for the massless fields of type {\bf N} and type {\bf D} solutions coincide when $\Lambda=0$. 

In this article we work with type {\bf D} and type {\bf N} solutions which are written a KS form, which means we can write them as
\begin{align*}
    d s^2=d\overline{s}^2+\phi \left( k_{\mu} dx^{\mu}\right)^{2}\,, \quad  g_{\mu\nu}= \overline{g}_{\mu\nu} + \kappa_{g}\,\phi_{g}\, k_{\mu}k_{\nu} \, .
\end{align*}
In this context, we have $\phi_{g}=\phi_{g_1}+\phi_{g_2}$, where $\phi_{g_1}$ and $\phi_{g_2}$ are the vacuum and matter contributions to the profile.

The other classical double copy mapping we mention is the Kerr-Schild double copy \cite{Monteiro:2014cda}, which states that certain classical GR solutions and gauge theory solutions are related: ${h_g}_{\mu\nu}$ is related to single copy field by the replacement of a Lorentz factor $k_{\mu}$ with a color factor $c^{a}$, resulting in the non-Abelian field ${A_{g}^{a}}_{\mu}=c^{a}\phi_{g}k_{\mu}$. The zeroth copy is obtained by replacing another $k_{\mu}$ with a color factor $c^{a'}$, resulting in the biadjoint scalar field $\Phi_{g}^{aa'}=c^{a}\widetilde{c}^{a}\phi_{g}$. 

The relation between the Kerr-Schild and the Weyl double copies have been studied, for example, in \cite{Caceres:2025eky}. The WDC double copy is consistent with the Kerr-Schild double copy \cite{Easson:2021asd,Easson:2022zoh}. The real part of $S_{(g_i)}$ is proportional to the scalar profile $\phi_{g_i}$ present in the Kerr-Schild form.  In what we present, the equations of motion in both classical double copies coincide. The Kerr-Schild form permits a straightforward decomposition of the Weyl spinor and we can state the results in a clear way. Even though the Kerr-Schild is connected to the Weyl double copy, our main interest is that, in bigravity, having a specific Kerr-Schild structure in both metrics generates an interaction which is linear in the profiles of the solutions. This permits the study of these kind of solutions, which will be mentioned in the next section.

\subsection{Examples of WDC in GR}

We focus on two type {\bf N} solutions: pp-waves, where the cosmological constant $\Lambda$ is zero, i.e., $\Lambda=0$, and Siklos-AdS waves, where $\Lambda$ is negative. We will check that these solutions lead to a Weyl double copy with equations of motion (\ref{typeN_eom_1}) for the fields. On the other hand, the type {\bf D} solutions we study are Kerr-Newman-(A)dS and the Plenba\'nski-Demia\'nski metric. Both solutions have a non-vanishing cosmological constant, which can be zero, positive, or negative. Having set the stage, we now explore certain solutions in GR and their Weyl double copy relations.

\subsubsection{pp-waves}
pp-waves is a type {\bf N} solution of GR equations in vacuum, which describes gravitational plane-fronted waves with parallel rays propagating in a Minkowski background. This solution was studied in the context of the Kerr-Schild double copy in \cite{Carrillo-Gonzalez:2017iyj}. In Poincar\'e coordinates, the metric of the pp-waves solution can be written in a Kerr-Schild form as
\begin{align}
    \nonumber d\overline{s}^2=\left(2dudv-dx^2- dy^2\right) \, , \quad k_{\mu}dx^{\mu}=-du \,, \quad \, \phi_{g}(u,x,y)=\frac{1}{\kappa_{g}}f_{2}(u) (x^{2}-y^{2}) \, ,
\end{align}
so that we have a `mostly-minus' Lorentzian signature. 
For this solution, the tetrad aligned to the corresponding principal null direction is
\begin{align}
\label{tetrad_pp}
    \left.e^a{}_{\mu}=\left(
\begin{array}
{cccc}\frac{1}{2}(1+\kappa_{g}\phi_{g}) & 1 & 0 & 0 \\
\frac{1}{2}(1-\kappa_{g}\phi_{g}) & -1 & 0 & 0 \\
0 & 0 & 1 & 0 \\
0 & 0 & 0 & 1
\end{array}\right.\right) \, . 
\end{align}
The flat limit of this metric is obtained by taking $f_{2}(u)=0$. 

For this type {\bf N} solution, the Weyl invariant and the Weyl spinor are
\begin{align}
    \psi_{4}=\psi_{g_1}=\frac{\kappa_{g}}{2} \left(\partial_{x} + i \,  \partial_{y}\right)^{2} \phi_{g_1}=2 f_{2}(u)\, ,\quad \Psi_{ABCD}= 2 f_{2}(u)\,o_{A}o_{B}o_{C}o_{D}\,.
\end{align}
Because we are in vacuum $\psi_{g_2}=0$ and  $\psi_{4}= \psi_{g_1}$.

In order to apply the double copy prescription (\ref{WDC_1}) we construct the gauge and scalar theory fields by means of (\ref{def_gauge_scalar_th}). We seek for $\Theta_{g_1}$, which will define the scalar theory and the gauge theory by means of $f^{(g_{1)}}_{AB}=\Theta_{g_{1}} o_{A} o_{B}$ for a type {\bf N} solution. We construct these fields using a function $\widetilde{\varphi}_{g_1}$, defined as $\widetilde{\varphi}_{g_1}\equiv\phi_{g_1}(f_{2}(u)\rightarrow \widetilde{\mathrm{f}}_{2}(u))$. $\Theta_{g_1}$ for this solution is defined using $\widetilde{\varphi}_{g_1}$, which is expressed as follows:
\begin{align*}
    \widetilde{\varphi}_{g_1}\equiv\widetilde{\varphi}_{g_1}(u,x,y)=\frac{1}{\kappa_{g}}\widetilde{\mathrm{f}}_{2}(u) (x^{2}- y^{2})\,, \qquad \Theta_{g_1}=-i \left(\partial_{x} + i \,  \partial_{y} \right)\widetilde{\varphi} \, .
\end{align*}
The idea is to associate the gravity theory solution $\phi_{g_1}$ with the function $\widetilde{\varphi}_{g_1}$, which in this case contains the arbitrary function $\widetilde{\mathrm{f}}_{2}(u)$. Then, we can identify $\Theta_{g_1}$ and therefore, the solution to the Maxwell equations, $f^{(g_1)}_{AB}$. Also, we can also extract $S_{(g_1)}$ from (\ref{def_gauge_scalar_th}), using $\Theta_{g_1} $ and $\psi_4$. Summarizing these results the single and zeroth copy field $f^{(g_1)}_{AB}$ and $S_{(g_1)}$ are 
\begin{align}
    \label{pp-waves-eom_GR}
    \nonumber &S_{(g_1)} =\frac{(\Theta_{g_1})^2}{\psi_{4}}=\frac{ \left[-i \left(\partial_{x} + i \,  \partial_{y}\right)\widetilde{\varphi}\right]^{2}}{\frac{\kappa_{g}}{2}\left(\partial_{x} + i \,  \partial_{y}\right)^{2}\phi}\,= -\frac{2}{{\kappa_{g}}^2}\, \frac{\widetilde{\mathrm{f}}_{2}(u)^{2}(x- iy)^{2}}{f_2(u)}\, , \\
    &f_{(g_1)\,A B}=\Theta_{g_1} o_{A}o_{B}= \,\left[-i\left(\partial_{x} + i \,  \partial_{y} \right)\widetilde{\varphi} \right]o_{A}o_{B} = -\frac{2}{\kappa_{g}}i\, \widetilde{\mathrm{f}}_{2}(u) (x-iy)\,o_{A}o_{B} \, .
\end{align}
The strength tensor $F^{(g_1)}_{\mu\nu}$ is obtained with (\ref{Fmunu_spinorial}) and $f^{(g_1)}_{AB}$.
Equivalently, we can construct $F^{(g_1)}_{\mu\nu}$ using the potential $A^{(g_1)}_{\mu}=\widetilde{\varphi}_{g_1} k_\mu$ as $F^{(g_1)}_{\mu\nu}\equiv \nabla_\mu A^{(g_1)}_{\nu}- \nabla_{\nu}A^{(g_1)}_{\mu}$. The resulting equations of motion for these fields (\ref{pp-waves-eom_GR}) are (\ref{typeN_eom_1}) in flat background, i.e., with $\Lambda=0$ are given by 
\begin{align}
    \nonumber \,^{(0)}\nabla_{\mu}F_{(g_1)}^{\mu\nu}=0\, , \qquad \,^{(0)}\Box\, S_{(g_1)}=0\,.
\end{align}
In this case, the Kerr-Schild is related to the real part of the Weyl double copy scalar as
\begin{align*}
    \nonumber \left. \mathfrak{Re}(S_{(g_1)}) \right|_{\widetilde{\mathrm{f}}_{2} \rightarrow f_2}  =-\frac{2}{\kappa_g}  \phi_{g_1}\,.
\end{align*}

\subsubsection{Siklos-AdS waves}

Siklos-(A)dS waves are a solution of GR for gravitational waves propagating in an AdS spacetime \cite{Stephani:2003tm,siklos1985lobatchevski}. Siklos waves in Poincaré coordinates permit a Kerr-Schild form, given by
\begin{align}
    \nonumber 
    d\overline{s}^2=\frac{l^2}{y^2}\left(2dudv-dx^2-dy^2\right), \\  
    k_\mu dx^\mu=\frac{l}{y}du\, ,\,\,\,  \phi_{g}=\frac{1}{\kappa_{g}}f_{3}(u)\left(\frac{y}{l}\right)^{3}, \ \ \,\,\, \Lambda=-\frac{3}{l^2} \,,
\end{align}
where $\phi_{g}=\phi_{g_1}$. The tetrad for this solution is
\begin{align}
\label{tetrad_Siklos}
    \left.e^a{}_{\mu}=\frac{l}{y}\left(
\begin{array}
{cccc}\frac{1}{2}(1+\kappa_{g}\phi_{g}) & 1 & 0 & 0 \\
\frac{1}{2}(1-\kappa_{g}\phi_{g}) & -1 & 0 & 0 \\
0 & 0 & 1 & 0 \\
0 & 0 & 0 & 1
\end{array}\right.\right).
\end{align}
For this vacuum solution,  $\psi_{4}=\psi_{g_1}$ and $\psi_{g_2}=0$, where
\begin{align}
    \nonumber \psi_{g_1}=\frac{\kappa_{g}}{2} \left(\frac{y}{l}\right)^{2}\left[\left(\partial_{x} + i \,  \partial_{y}\right)^{2}\right]\phi_{g_1}=\Lambda \,\kappa_g \,\phi_{g_1}\, , \quad \Psi_{ABCD}=\left(\Lambda \,\kappa_g \,\phi_{g_1} \right)\, o_{A}o_{B}o_{C}o_{D}\, .
\end{align}
In order to find the Weyl double copy fields, $\phi_{g_1}$ is associated again with $\widetilde{\varphi}_{g_1}$, where we use an arbitrary function $\widetilde{\mathrm{f}}_{3}(u)$. We define also $\Theta_{g_1}$, so that we have
\begin{align*}
    \widetilde{\varphi}_{g_{1}}\equiv\widetilde{\varphi}_{g_1}(u,x,y)=\frac{1}{\kappa_{g}}\widetilde{\mathrm{f}}_{3}(u) \left(\frac{y}{l}\right)^{3}\,, \qquad \Theta_{g_1}=\frac{2}{3}i\left(\frac{y}{l}\right)\left[\left(\partial_{x} + i \,  \partial_{y} \right)\widetilde{\varphi}_{g_1} \right] \, .
\end{align*}
Using (\ref{def_gauge_scalar_th}), the respective field $f^{(g_1)}_{A B}$ and scalar function $S_{(g_1)}$ are written as
\begin{align}
    \nonumber S_{(g_1)} =\frac{(\Theta_{g_1})^{2}}{\psi_{4}}=-\frac{4}{3\, \kappa_{g}}\,\frac{\widetilde{\mathrm{f}}_{3}(u)}{f_3(u)}\,\widetilde{\varphi}_{g_1}\, , \qquad 
    f^{(g_1)}_{A B}= \Theta_{g_1}o_{A}o_{B} = \left(-\frac{2}{l}\,\widetilde{\varphi}_{g_1}\right)\,o_{A}o_{B}\,.
\end{align}
The fields obey (\ref{typeN_eom_1}), where $\Lambda$ is non-vanishing and we have a similar relation between $S_{(g_1)}$ and $\phi_{g_1}$ as in the previous example. A similar relation will also occur when coupling this solution to external matter.

\subsubsection{Siklos-Maxwell waves}
We can couple Siklos-(A)dS waves to a Maxwell field \cite{Stephani:2003tm,siklos1985lobatchevski}. The electromagnetic energy-momentum tensor is defined using ${A_{M}}_{\mu}$, given by $A_{M\,\mu}=(A_{M}^{(u)},0,0,0) \,$, where $A_{M}^{(u)}$ solves the harmonic equation. The solution is the real part of a holomorphic function on the complex wavefront coordinate $z=x+iy$,
$A_{M}^{(u)}(u,x,y)=\partial_{z} a(u,z)+\overline{\partial_{z}a(u,z)}\,$, where $a(u,z)=\frac{1}{2}D_{2}(u)z^{2}$. The Kerr-Schild profile for this solution is $\phi_g=\phi_{g_1}+\phi_{g_2}$, where $\phi_{g_1}$ is the same as before and the profile which accounts for the matter content of the solution is $\phi_{g_2}$, defined as
\begin{align*}
    \phi_{g_2}=\Delta_{g} D_{2}(u)^2\,y^4\,, \quad  \Delta_{g}\equiv\frac{\kappa_{g}}{4\,\pi\,l^2}\,,\quad m^{2}_{\text{er}}=\frac{4}{l^2}=\frac{2}{3}\left(\rho_{(g_2)}^{\text{grav}}- \Lambda\right)\,,
\end{align*}
where we have used gravitational energy density of the solution $ \rho_{(g_2)}^{\text{grav}}\equiv R^{u}{}_{u}$. We have also defined the square of the resonance mass, which we will use later. The Weyl invariant corresponding to this contribution is written as
\begin{align*}
\psi_{g_2}=\frac{\kappa_g}{2} \left(\frac{y}{l}\right)^{2}\left[\left(\partial_{x} + i \,  \partial_{y}\right)^{2}\right]\phi_{g_2}(u,x,y) \,=2\,\Lambda\, \kappa_{g}\,\phi_{g_2}.
\end{align*}
 Thus we define $\widetilde{\varphi}_{g_2}$ and $\Theta_{g_2}$ as
 \begin{align*}
     \widetilde{\varphi}_{g_2}=\Delta_{g} \widetilde{D}_{2}(u)^2\,y^4\,, \quad \Theta_{g_2}=-\frac{3}{l}\,\widetilde{\varphi}_{g_2},\,
 \end{align*}
where we have used an arbitrary function $\widetilde{D}_{2}(u)$. We obtain that the scalar function and spinor strength as before, leading to
\begin{align}
    \nonumber 
    S_{(g_2)}=-\frac{3}{2\,\kappa_g}\frac{\widetilde{D}_2(u)^2}{D_2(u)} \widetilde{\varphi}_{g_2}\,, \qquad f^{(g_2)}_{AB}=\left(-\frac{3}{l}\,\widetilde{\varphi}_{g_{2}}\right)\,o_A o_B.
\end{align}
Then, the equations of motion for the scalar and gauge field can be written as
\begin{align*}
    \overline{\nabla}_{\mu}F_{(g_2)}^{\mu\nu}- \frac{\overline{R}}{6}\,A^{\nu}_{(g_{2})}+\frac{2}{3}\left(\rho_{(g_2)}^{\text{grav}}- \Lambda\right)A^{\nu}_{(g_{2})} &=0\, , \\
    \overline{\Box}\,S_{(g_2)} +\frac{2}{3}\left(\rho_{(g_2)}^{\text{grav}}- \Lambda\right)S_{(g_2)}&=0\,.
\end{align*}
This result is similar to \cite{Han:2022mze}, but in this case we consider a type {\bf N} solution. We note that these equations can be written as in (\ref{typeN_eom_1}), where we have a Proca equation with a term proportional to the scalar curvature and a massive Klein-Gordon equation, both with mass $m_{\text{res}}=m_{\text{er}}$ in an AdS background.

\subsubsection{Siklos-Scalar waves}
We can couple Siklos-(A)dS waves to a scalar matterless field $\Phi$ \cite{Stephani:2003tm,siklos1985lobatchevski}. The scalar field $\Phi$ obeys the wave equation, so that the field is an arbitrary function of the retarded time $\Phi = \Phi (u)$. The profile $\phi_{g_2}$ which accounts for the scalar matter content of the solution is given by
\begin{align*}
    \phi_{g_2}=\Delta_{g} \dot{\Phi}^2\,y^2\,, \quad  \Delta_{g}\equiv\frac{\kappa_{g}}{2}\,,\qquad m^{2}_{\text{sr}}=-\frac{2}{l^2}=-\frac{\overline{R}_{g}}{6}=-\frac{1}{3}\left(\rho_{(g_2)}^{\text{grav}}- \Lambda\right)\,,
\end{align*}
and we have also defined the square of the resonance mass. The Weyl invariant corresponding to this contribution is written as
\begin{align*}
\psi_{g_2}=\frac{\kappa_g}{2} \left(\frac{y}{l}\right)^{2}\left[\left(\partial_{x} + i \,  \partial_{y}\right)^{2}\right]\phi_{g_2}(u,x,y) \,=\frac{1}{3}\,\Lambda\, \kappa_{g}\,\phi_{g_2}.
\end{align*}
Now  we define $\widetilde{\varphi}_{g_2}$ and $\Theta_{g_2}$ as
 \begin{align*}
     \widetilde{\varphi}_{g_2}=\Delta_{g} \dot{\widetilde{\Phi}}(u)^2\,y^2\,, \ \ \ \ \ \quad \Theta_{g_2}=-\frac{1}{l}\,\widetilde{\varphi}_{g_2},\,
 \end{align*}
where we employed a function $\widetilde{\Phi}(u)$. The scalar function and spinor strength are
\begin{align}
    S_{(g_2)}=-\frac{1}{\kappa_g}\frac{\dot{\widetilde{\Phi}}(u)^2}{\dot{\Phi}(u)^2} \widetilde{\varphi}_{g_2}\,, \qquad f^{(g_2)}_{AB}=\left(-\frac{1}{l}\,\widetilde{\varphi}_{g_{2}}\right)\,o_A o_B.
\end{align}
The equations of motion for the matter fields are in (\ref{typeN_eom_1}), with $m_\text{res}=m_{\text{sr}}$. 

We note that these matter fields are mapped to massless equations, namely Maxwell and Klein-Gordon with a conformal term. This can be seen from (\ref{typeN_bg_2_eom}), where the value of $m_{\text{res}}$ reduces the conformal Proca equation to a Maxwell equation, and the massive Klein-Gordon becomes massless with a conformal curvature term.

\subsubsection{Dyonic Kerr-Newman-(A)dS}
The Weyl double copy of Kerr solution was studied in \cite{Luna:2018dpt} and the charged case in \cite{Easson:2022zoh,Easson:2021asd}. Kerr-(A)dS solution was also studied in the context of the Weyl double copy in \cite{Han:2022mze}. These solutions permit a Kerr-Schild form when written in ellipsoidal coordinates \cite{Gibbons:2004uw}. Here we consider the Weyl double copy of the dyonic Kerr-Newman-(A)dS solution in its Kerr-Schild form \cite{Ayon-Beato:2025ahb}.

The dyonic Kerr-Newman-(A)dS is solution of GR which describes a black hole with mass $m$, rotation parameter $a$, electric charge $e$ and magnetic charge $p$, which lives in a maximally symmetric background. It was derived by using the Kerr-Schild ansatz in \cite{Ayon-Beato:2025ahb} and can be written as
\begin{align}
    \nonumber d\bar{s}^{2}=\frac{\Delta}{\Omega}\, \xi \,  dt^{2}-\frac{\Sigma \,  dr^{2}}{(r^{2}+a^{2})\, \xi}-\frac{\Sigma \, d\theta^{2}}{\Delta}-\frac{(r^{2}+a^{2})}{\Omega}\sin^{2}\theta d\varphi^{2}\, , 
\end{align}
where
\begin{align}
    \nonumber \xi \equiv1-\frac{\Lambda r^2}{3},\quad\Sigma\equiv r^2+a^2\cos^2\theta,\quad\Delta\equiv1+\frac{\Lambda}{3}a^2\cos^2\theta,\quad\Omega\equiv1+\frac{\Lambda}{3}a^2\,. 
\end{align}
The solution is determined by
\begin{align}
\nonumber \phi_{g} &=-\frac{\kappa_{g}}{2}\frac{[2mr-(e^2+p^2)]}{8\,\pi\,\Sigma} \, , \quad {A_g}_{\mu}\equiv \frac{e\,r}{\Sigma}k_{\mu}- \frac{p \cos\theta}{\Sigma} k^*_\mu\,,\\
k_{\mu}&=\left(\frac{\Delta}{\Omega}, \frac{\Sigma}{(r^{2}+a^{2})\,\xi},0,-\frac{a \sin^{2}\theta}{\Omega}\right) \,, \quad k^*_\mu=\left(\frac{\xi}{\Omega} a,0,0,-\frac{r^2+a^2}{\Omega}\right),
\end{align}
where ${A_{g}}_{\mu}$ is the vector potential of the electromagnetic field strength. The tetrad aligned with the principal null directions of the solution is written as
{\begin{align}
\label{tetrad_dKNg}
    e^a{}_{\mu}=
\begin{pmatrix}
    \dfrac{\Delta}{\Omega}\sqrt{\dfrac{\Delta_{g_r}}{\Sigma}} & \sqrt{\dfrac{\Sigma}{\Delta_{g_r}}}\dfrac{\Sigma\,\kappa_{g}\, \phi_g }{\left(r^2+a^2\right)\,\xi} & 0 & -\sqrt{\dfrac{\Delta_{g_r}}{\Sigma}}\dfrac{a \sin^2 \theta}{\Omega}  \\
    0 & \dfrac{\Sigma}{\Delta_{g_r}} & 0 & 0 \\
    0 & 0 & \sqrt{\dfrac{\Sigma}{\Delta}} & 0 \\
    -\sqrt{\dfrac{\Delta}{\Sigma}} \dfrac{\xi a \sin \theta}{\Omega} \, & 0 & 0 & \sqrt{\dfrac{\Delta}{\Sigma}} \dfrac{\sin \theta \, \left(r^{2}+a^2\right)}{\Omega}
\end{pmatrix}\, ,
\end{align}}
where $\Delta_{g_r}=(r^2+a^2)\,\xi+\kappa_{g}\phi_g \,\Sigma$. The Weyl invariant $\psi_4$ is the sum of the terms
\begin{align}
\psi_{g_1}=\frac{{\kappa_{g}}^2}{16\pi}\frac{m}{\left(r-i a \cos\theta\right)^{3}}\, ,\quad \psi_{g_2}=-\frac{{\kappa_{g}}^2}{16\pi}\frac{e^{2}+p^2}{\left(r-i a \cos\theta\right)^{3}\left(r+i a \cos\theta\right)}\,,
\end{align}
which applies for both cases, $\Lambda=0$ and $\Lambda\ne 0$. 
The self-dual part of the Weyl spinor is $\Psi_{ABCD}=(\psi_{g_1}+\psi_{g_2})\,o_{(A}o_{B}l_{C}l_{D)}$. By mapping as for type {\bf N} solutions, we associate the parameters of the gravity theory, $m,e$ and $p$ with the parameters of the gauge theory $\widetilde{m},\widetilde{e},\widetilde{p}$. Then we have 
\begin{align}
    \nonumber S_{(g_1)} =\frac{1}{6}\frac{16 \pi}{\kappa_{g}^{2}}\frac{\widetilde{m}^{2}}{m\,\left(r-i a \cos\theta\right)} \, , \quad
    f^{(g_1)}_{A B}= -\frac{\widetilde{m}}{\left(r-i a \cos\theta\right)^{2}}\,o_{(A}l_{B)} \, , \quad A^{(g_1)}_{\mu}=-\frac{\widetilde{m}r}{\Sigma}\,k_\mu\, 
\end{align}
and
\begin{align}
    \nonumber S_{(g_2)} &=-\frac{1}{6}\frac{16 \pi}{\kappa_{g}^{2}}\frac{(\widetilde{e}^{2}+\widetilde{p}^{2})^2}{(e^{2}+p^2)\,\left(r-ia\cos\theta\right)\left(r+ia\cos\theta\right)} \, , \\
    \nonumber f^{(g_2)}_{A B}&= \,\frac{\widetilde{e}^{2}+\widetilde{p}^{2}}{\left(r-ia\cos\theta\right)^{2}\left(r+ia\cos\theta\right)}\,o_{(A}l_{B)}\,.
\end{align}
Then we define
\begin{align}
    \nonumber \left.\rho_{(g_2)}^{\text{grav}}\, \right|_{\Lambda =0}\equiv \left.R^{t}{}_{t}\right|_{\Lambda =0} = \,^{(0)}\rho_{(g_2)}^{\text{grav}}\equiv \,^{(0)}R^{t}{}_{t}=\frac{\kappa_{g}^{2}}{16 \pi}\frac{(e^{2}+p^2)\, (r^2-a^2 \cos^2 \theta +2 a^2)} {\left(r^2+a^2 \cos^2\theta\right)^{3}} \,,
\end{align}
and we can define also the sources
{\begin{align}
    \nonumber \,^{(0)}{J}^{\nu}_{(g_2)}&=\frac{(\widetilde{e}^{2}+\widetilde{p}^{2})}{\left(r^2+a^2\cos^2\theta\right)^{3}}\left[(r^2-a^2\cos^2\theta+2a^2 )\,\delta^{\nu}{}_{t}+2 a\,\delta^{\nu}{}_{\phi}\right]\, , \\ 
    \nonumber \widetilde{\Lambda}_{(g_2)}^{\nu}&= -\frac{1}{3}\frac{(r^2-a^2 \cos^2\theta)}{(r^2+a^2\cos^2\theta)^3} \,\Lambda\, a \, (\widetilde{e}^{2}+\widetilde{p}^{2})\,\delta^{\nu}_{\phi}\,,\\
\nonumber \,^{(0)}j_{(g_2)}&=\frac{4}{3}\frac{64\pi^2}{\kappa_{g}^4}\frac{(\widetilde{e}^{2}+\widetilde{p}^{2})^2}{(e^{2}+p^2)^2}\,\,^{(0)}\rho_{(g_2)}^{\text{grav}}  \,,\\
    \widetilde{\lambda}_{(g_2)}&=
    -\frac{4}{3}\frac{64\pi^2}{\kappa_{g}^4}\frac{(\widetilde{e}^{2}+\widetilde{p}^{2})^2}{(e^{2}+p^2)^2}\, \,^{(0)}\rho^{\text{grav}}_{(g_2)} \left(\frac{1}{3}\frac{(r^2-a^2\cos^2{\theta}- 2\,r^2\cos^2\theta)}{r^2-a^2\cos^2\theta +2 a^2}\,\Lambda\, a^2\right).\,
\end{align}
}The fields defined obey equations (\ref{typeD_eom_1}) with their respective sources. Taking the limit $\Lambda=0$, we obtain equations (\ref{typeD_eom_1}) in a flat background
\begin{align}
\label{KN(A)dS_eom_flat_1} 
    \,^{(0)}\nabla_{\mu}F_{(g_1)}^{\mu\nu}&=0\, , \qquad &&\,^{(0)}\Box\, S_{(g_1)}=0\,,\\
    \,^{(0)}\nabla_{\mu}F_{(g_2)}^{\mu\nu}&=\,^{(0)}J^{\nu}_{(g_2)}\, , \quad &&\,^{(0)}\Box\, S_{(g_2)}= \,^{(0)}j_{(g_2)} \,.
\end{align}

Focusing on the scalar theory, we could factorize the equation of motion by writing
\begin{align*}
    V(r,\theta)\equiv-\left(\frac{32\,\pi}{{\kappa_{g}}^2}\right)\frac{(r^2+a^2\cos^2\theta)}{e^2+p^2}\, \,^{(0)}\rho_{(g_2)}^{\text{grav}} \left(1-\frac{1}{3}\frac{(r^2-a^2\cos^2{\theta}- 2\,r^2 \cos^2\theta)}{r^2-a^2\cos^2\theta +2 a^2}\,\Lambda\, a^2\right),\,
\end{align*}
which permit us to rewrite the external sources as so that
\begin{align}
    \overline{\Box}\,S_{(g_2)} - \frac{\overline{R}}{6}\,S_{(g_2)} - V(r,\theta)\, S_{(g_2)}=0\,, \quad \,^{(0)}j_{g_2}+\widetilde{\lambda}_{g_2} = V(r,\theta)\, S_{(g_2)}\,.
\end{align}
In this term, $V(r,\theta)$ depends on the coordinates. Therefore, this contribution is interpreted as an external source in the scalar theory and gauge theory.

\subsubsection{Pleba\'nski-Demia\'nski solution}
Pleba\'nski-Demia\'nski solutions \cite{Plebanski:1975xfb} permit a double Kerr-Schild form,
\begin{align*}
    d s^2=d\overline{s}^2+\phi_g \left( l_{\mu} dx^{\mu}\right)^{2}+\chi_g \left(k_{\mu} dx^{\mu}\right)^{2}\,, \quad g_{\mu\nu}= \overline{g}_{\mu\nu}+h_{\mu\nu} \, ,  \quad h_{\mu\nu} = \phi_{g} l_{\mu}l_{\nu} +\chi_g k_{\mu}k_{\nu} \, .
\end{align*}
We can write down the vacuum Pleba\'nski-Demia\'nski metric in double KS coordinates $(\widetilde{\tau},\widetilde{\sigma},p,q)$ \cite{Chong:2004hw}, where the background line element is defined as
{\small
\begin{align}
    \label{Pleb_bg}
    \nonumber d\overline{s}^2&=\frac{1}{q^2-p^2}\left[\overline{\Delta}_p(d\widetilde{\tau}+q^2d\widetilde{\sigma})^2+\overline{\Delta}_q(d\widetilde{\tau}+p^2d\widetilde{\sigma})^2\right]+2(d\widetilde{\tau}+q^2d\widetilde{\sigma})dp+2(d\widetilde{\tau}+p^2d\sigma)dq\,, 
\end{align}}
where
\begin{align}
    \overline{\Delta}_p&\equiv\gamma-\varepsilon\,p^2+\frac{\Lambda}{3} p^4,\quad\overline{\Delta}_q\equiv-\gamma+\varepsilon\, q^2-\frac{\Lambda}{3} q^4\, .
\end{align}
This solution is coupled to an electromagnetic field ${A_{g}}_{\mu}$, and the scalar function and null vector are given by
\begin{align}
    \label{null vects_gr}
     \nonumber \phi_g&=-\frac{\kappa_{g}}{2}\frac{2mq-e^2}{8 \,\pi \,(q^2-p^2)} \, , \,\, l_\mu=(1,p^2,0,0) \,,\, \chi_g=-\frac{\kappa_{g}}{2}\frac{2np+\mathrm{g}^{2}}{8 \, \pi \, (q^2-p^2)}\, , \,\,  k_\mu=(1,q^2,0,0)\, ,\\
     {A_{g}}_{\mu}&=\frac{1}{q^2-p^2}\left(e\, q \,l_{\mu}+\mathrm{g}\,p\,k_{\mu}\right)\,.
\end{align}
This is the most general electrovacuum type {\bf D} solution and allows a double Kerr-Schild form after an analytical continuation to a $2+2$ metric \cite{Plebanski:1975xfb,Chong:2004hw}. It consists on seven parameters $m,n, e, \mathrm{g}, \varepsilon, \gamma$ and $\Lambda$, which can be interpreted as the the mass, NUT parameter, electric and magnetic charge,  kinematic parameters and cosmological constant when considering specific limits and transformations of the metric.

In these coordinates, we use $\eta_{ab}=\text{diag}(1,1,-1,-1)$ and the tetrad for this solution is given by
\begin{align}
\label{tetrad_PDg}
    e^a{}_{\mu}=\frac{1}{\sqrt{2}}
\begin{pmatrix}
1+\Delta_q & p^2(1+\Delta_p) & 0 & 1 \\
i\,(1-\Delta_q) & i p^2(1-\Delta_q) & 0 & -i \\
-(1+\Delta_p) & -q^2(1+\Delta_p) & 1 & 0 \\
-i(1-\Delta_p) & -iq^2(1-\Delta_p) & -i & 0
\end{pmatrix},\, 
\end{align}
where
\begin{align}
    \Delta_{g_p}=\frac{1}{2}\left(\frac{\overline{\Delta}_p}{p^2-q^2}- \kappa_g\,\chi_{g}\right),\quad \ \ \ \ \ \Delta_{g_q}=-\frac{1}{2}\left(\frac{\overline{\Delta}_q}{p^2-q^2}-\kappa_{g}\,\phi_g\right). \,
\end{align}
For both cases $\Lambda=0$ and $\Lambda\ne 0$, the Weyl invariant $\psi_4$ is the sum of the terms
\begin{align}
    \psi_{g_1}=-\frac{{\kappa_{g}}^2}{16\,\pi}\frac{(m+n)}{\left(p-q\right)^{3}}\, ,\quad \psi_{g_2}=\frac{{\kappa_{g}}^2}{16\,\pi}\frac{(e^{2}-\mathrm{g}^{2})}{\left(p-q\right)^{3}\left(p+q\right)}\,,
\end{align}
so that the self-dual part of the Weyl spinor is $ \Psi_{ABCD}=(\psi_{g_1}+\psi_{g_2})\,o_{(A}o_{B}l_{C}l_{D)}$. Considering an analogous mapping as for type {\bf N} solutions, we can associate the parameters of the gravity theory, $m,n,e,\mathrm{g}$ with the parameters of the gauge theory $\widetilde{m},\widetilde{n},\widetilde{e},\widetilde{\mathrm{g}}$. Then 
\begin{align}
    \nonumber S_{(g_1)} &=-\frac{2}{3}\frac{16\,\pi}{{\kappa_{g}}^2}\frac{\left(\widetilde{m}+\widetilde{n}\right)^{2}}{\left(m+n\right)\,\left(p-q\right)} \, , \quad
    &&f^{(g_1)}_{A B}= \left(-2\, \frac{\widetilde{m}+\widetilde{n}}{\left(p-q\right)^{2}}\right)\,o_{(A}l_{B)} \, , \\
    \nonumber S_{(g_2)} &=\frac{2}{3}\frac{16\,\pi}{{\kappa_{g}}^2}\frac{\left(\widetilde{e}^{2}-\widetilde{\mathrm{g}}^{2}\right)^{2}}{\left(e^{2}-\mathrm{g}^{2}\right)\,\left(p-q\right)\left(p+q\right)} \, , \quad
    &&f^{(g_2)}_{A B}= \left(2 \,\frac{\widetilde{e}^{2}-\widetilde{\mathrm{g}}^{2}}{\left(p-q\right)^{2}\left(p+q\right)}\right)\,o_{(A}l_{B)},
\end{align}
where
\begin{align}
    \nonumber 
    A^{(g_1)}_{\mu}=-\frac{\widetilde{m}+\widetilde{n}}{q^2-p^2}\left( q \,l_{\mu}+,p\,k_{\mu}\right)\,.
\end{align}
The real part of $S_{(G_i)}$ is proportional to the Kerr-Schild profile $\phi_{G_i}$. When the cosmological constant is non-zero, the equations of motion of the single copy fields, $F_{\mu\nu}^{(g_1)},$ $F_{\mu\nu}^{(g_2)}$ are the same equations as in  (\ref{typeD_eom_1}) with the sources given by the following relations:
\begin{align}
    \nonumber \,^{(0)}J^{\nu}_{(g_2)}&\equiv\frac{(\widetilde{e}^{2}-\widetilde{\mathrm{g}}^{2})}{\left(p^{2}-q^{2}\right)^{3}}\left[(p^{2}+q^{2})\delta^{\nu}{}_{\tau}-2\delta^{\nu}{}_{\sigma}\right],\,  \\
    \nonumber \widetilde{\Lambda}_{(g_2)}^{\nu}&=0,\,\\
    \nonumber \,^{(0)}j_{(g_2)}&\equiv\frac{4}{3}\frac{16\,\pi}{{\kappa_{g}}^2}\frac{\left(\widetilde{e}^{2}-\widetilde{\mathrm{g}}^{2}\right)^{2}}{\left(e^{2}-\mathrm{g}^{2}\right)^{2}}\left[\varepsilon\,\,^{(0)}\rho_{(g_2)}^{\text{grav}}+2\gamma_{p} \,\frac{(e^{2}-\mathrm{g}^{2})}{\left(q+p\right)^{3}\left(p-q\right)^{3}}\right],\,\\
    \lambda_{(g_2)}&\equiv
    -\frac{8}{9}\left(\frac{16\,\pi}{{\kappa_{g}}^2}\right)^2\frac{\left(\widetilde{e}^{2}-\widetilde{\mathrm{g}}^{2}\right)^{2}}{\left(e^{2}-\mathrm{g}^{2}\right)^{2}}\,\,^{(0)}\rho_{(g_2)}^{\text{grav}}\,\left(\frac{p^{2}q^{2}}{p^{2}+q^{2}}\, \Lambda\right),\,
\end{align}
where
\begin{align}
    \nonumber \,^{(0)}\rho_{(g_2)}^{\text{grav}}=\,^{(0)}R^{\tau}{}_{\tau}=-\frac{{\kappa_{g}}^2}{16\,\pi}\frac{(e^{2}-\mathrm{g}^{2})\, (p^{2}+q^{2})} {\left(q+p\right)^{3}\left(p-q\right)^{3}} \,=\rho_{(g_2)}^{\text{grav}}+\Lambda\,.
\end{align}
This is consistent with \cite{Easson:2021asd} when $\Lambda=0$, where the prescription the authors provided constrained $\gamma_{p}=0$. In such case, we have equations (\ref{typeD_eom_1}) in a flat background. Also, in the last equality, we note that $\,^{(0)}\rho_{(g_2)}^{\text{grav}}$ relates to $\rho_{(g_2)}^{\text{grav}}$ in the same way as in Siklos solution and which does not happen in Kerr-Newman-(A)dS solution.

Similar to Kerr-Newman-(A)dS, by setting $\gamma_{p}=0$ we can factorize the equation of motion by substituting the sources with a product of a potential and a field and interpret such term as an external source in the scalar theory.

\medskip
\section{Weyl double copy in bigravity}\label{Sec_WDC in BG}
\subsection{Bigravity theory}

The theory of bigravity of Hassan and Rosen \cite{Hassan:2011zd} is a ghost-free modification of GR at low energies. It consists of the usual spin-2 massless degree of freedom from GR and an additional spin-2 massive field. The interaction between the two fields is defined by a non-derivative potential $\mathcal{U}[g,f]$ that depends on an interaction matrix $\gamma^{\mu}{}_{\nu}$. The action for the theory of bigravity in the vacuum case is given by
{\small\begin{align}
    \label{action_big}
    S_{bi}[g, f]& =  \frac{1}{{\kappa_{g}}^2} \int d^{4} x \sqrt{-g} \, R[g]+\frac{1}{{\kappa_{f}}^2} \int d^{4} x \sqrt{-f} \mathcal{R}[f] -\frac{2\,m^{2}}{\kappa^2} \int d^{4} x \sqrt{-g} \, \mathcal{U}[g, f]\,,\\
    \mathcal{U}[g,f] & =\sum_{k=0}^{4} b_{k} \mathcal{U}_{k}(\gamma) \, , \qquad \gamma^{\mu}{}_{\nu}=\sqrt{g^{\mu\alpha}f_{\nu\alpha}}, \, 
\end{align}
where ${\kappa_{g}}^{2}=16\pi G_{g}$ and ${\kappa_{f}}^{2}=16\pi G_{f}$ are the gravitational couplings for each of the metrics and $\kappa^2$ is a function of both couplings. The parameter $m$ is the mass is related to the massive graviton. 
 
By contemplating matter content in the theory, the action is modified to
$S= \, S_{\text{bi}}+S_{\text{Mat}}$, where $S_{\text{Mat}}$ accounts for the matter content and we have different options for the coupling of matter. One option is to couple matter independently to the  metrics, i.e., coupling matter content to the metric $g_{\mu\nu}$, which leads to an energy-momentum tensor ${T_{g}}_{\mu\nu}$ and an independent matter content to the metric $f_{\mu\nu}$ which results in ${T_{f}}_{\mu\nu}$, both tensors defined as the variation of the corresponding matter Lagrangian $\mathcal{L}^{G}_{\text{M}}$.

We can also couple matter to both metrics in a different way by using the effective metric ${g_{E}}_{\,\mu\nu}$ \cite{deRham:2014naa} constructed with $g_{\mu\nu}$ and $f_{\mu\nu}$. Then, ${T_{G}}_{\mu\nu}$ are related to the energy-momentum tensor constructed with the effective metric, ${T_E}_{\mu\nu}$. We will work with a specific Kerr-Schild ansatz in bigravity, where the energy-momentum tensors of both metrics are related as
\begin{align}
    \label{eff_metric}
    \frac{1}{\alpha}{T_{g}}_{\mu\nu}=\frac{C}{\beta}{T_{f}}_{\mu\nu}=(\alpha+\beta\,C){T_{E}}_{\mu\nu}\,,\qquad {g_{E}}_{\,\mu\nu}\equiv \alpha^{2}g_{\mu\nu}+2\alpha\beta g_{\mu\lambda}\,\gamma^{\lambda}{}_{\nu}+\beta^2f_{\mu\nu},\, 
\end{align}
where we have also defined the effective metric ${g_E}_{\mu\nu}$ using the constants $\alpha,\beta$. 

We will consider type {\bf D} and use matter independently coupled, whereas for type {\bf N} solutions we use the effective metric to couple external matter. In both cases, the equations of motion for the action in (\ref{action_big}) can be written as:
\begin{align}
\label{eom_big}
    \nonumber {G_{g}}^{\mu}{ }_{\nu}&= \, {Q_{g}}^{\mu}{}_{\nu} + \frac{{\kappa_{g}}^2}{2}{T_{g}}^{\mu}{}_{\nu}\,, \, &&{Q_{g}}^{\mu}{}_{\nu} \equiv \frac{2\,\kappa^{2}}{m^2\,{\kappa_{g}}^2}\frac{g^{\mu \alpha}}{\sqrt{-g}} \frac{\delta(\sqrt{-g} \, \mathcal{U})}{\delta g^{\alpha \nu}}, \\
    {G_{f}}^{\mu}{}_{\nu} &=  {Q_{f}}^{\mu}{}_{\nu} +\frac{{\kappa_{f}}^2}{2}{T_{f}}^{\mu}{}_{\nu}\, ,\, && {{Q}_{f}}^{\mu}{}_{\nu}\equiv \frac{2\,\kappa^{2}}{m^2\,{\kappa_{g}}^2} \frac{ f^{\mu \alpha}}{\sqrt{-f}} \frac{\delta(\sqrt{-g} \, \mathcal{U})}{\delta f^{\alpha \nu}}\,.
\end{align}
Here ${G_{g}}^{\mu}{}_{\nu}$ and ${G_{f}}^{\mu}{}_{\nu}$ are the Einstein tensors for the metrics $g_{\mu\nu}$ and $f_{\mu\nu}$ respectively. The interaction tensor ${Q_{G}}^{\mu}{}_{\nu}$ contains the information about the interaction between the metrics.

The form of the interaction in (\ref{action_big}) can be complex because of its square root structure. As commented before, we work with a family Kerr-Schild family solutions in bigravity, which allows a closed form for the interaction between the metrics $\mathcal{U}[g,f]$ \cite{Ayon-Beato:2015qtt}. With this ansatz, the metrics take the form
\begin{align}
    \label{ks_ansatz_bg}
    \nonumber \quad d s_{g}^2&=d\overline{s}_{g}^2+\phi_{g}\left( k_{\mu} dx^{\mu}\right)^{2}\,, \quad  &&g_{\mu\nu}=\overline{g}_{\mu\nu}+\kappa_{g}h_{\mu\nu} \, ,  \quad h_{\mu\nu} = \phi_{g}k_{\mu}k_{\nu} \, , \\
    d s_{f}^2&=C^2\left(d\overline{s}_{g}^2+\phi_{f} \left( k_{\mu} dx^{\mu}\right)^{2}\right)\,, \quad &&f_{\mu\nu}\, = \, C^{2}(\overline{g}_{\mu\nu}+\kappa_{f}\mathscr{h}_{\mu\nu}) \, , \quad \mathscr{h}_{\mu\nu} = \phi_{f}k_{\mu}k_{\nu} \,.
\end{align}
The profile in this case are decomposed as $\phi_{G}=\phi_{G_1}+\phi_{G_2}+\sum\phi_{G_j}$, where $\phi_{G_1}$ and $\phi_{G_2}$ are  the vacuum and matter profiles, and the index $j$ in $\phi_{G_j}$ stands for the massive modes that can be present.
The interaction tensors with this KS ansatz are given by
\begin{align}
    \nonumber Q^{\mu}{}_{\nu} &= \kappa_{g}\left[A_{g} \delta^{\mu}{}_{\nu} - B_{g} \left(\kappa_{g}h^{\mu}{}_{\nu}-C^{2}\kappa_{f}\mathscr{h}^{\mu}{}_{\nu}\right)\right]  \, , \\
    {{Q}_{f}}^{\mu}{}_{\nu} &= C^{2}\kappa_{f}\left[\,{A}_{f}\delta^{\mu}{}_{\nu} + B_{f} \left(\kappa_{g}h^{\mu}{}_{\nu}-C^{2}\kappa_{f}\mathscr{h}^{\mu}{}_{\nu}\right)\right]\, , 
\end{align}
where we have defined
\begin{align}
    \nonumber A_{g} \equiv \frac{m^{2} \kappa_{g}}{\kappa^{2}}(P_{1}) \, , \quad B_{g} \equiv \frac{m^{2} \kappa_{f}}{\kappa^{2}}\left(\frac{P_{2}}{C^{5}}\right) \, , \quad A_{f} \equiv \frac{m^{2} \kappa_{g}}{\kappa^{2}}(C P_{0}) \, , \quad  B_{g} \equiv \frac{m^{2} \kappa_{f}}{\kappa^{2}}\left(\frac{P_{0}}{C^{5}}\right) \, . 
\end{align}
Here, $P_{i}$ depends on the coupling constants $b_{k}$. For backgrounds with constant curvature, we can further set the value of $A_{G}$ in order to cancel the diagonal contributions. For the type {\bf D} solutions we study, $B_{G}=0$ and for type {\bf N} we have
\begin{align*}
    B_{g}=\frac{C^{2}\kappa_{g}}{2(C^{2}\kappa_{g}{}^{2}+\kappa_{f}{}^{2})}\widehat{m}^{2}=\frac{C^{6}\kappa_{g}}{\kappa_{f}}B_{f}\,.
\end{align*}    
By writing the equations (\ref{eom_big}) in terms of the deviations $h_{\mu\nu}$ and $\mathscr{h}_{\mu\nu}$, we obtain
\begin{align}
    \label{eom_bg_spin2}
    \nonumber \,^{(g)}\overline{\mathcal{E}}(h^{\mu}{}_{\nu}) -\frac{\overline{R}_{g}}{4} h^{\mu}{}_{\nu} = &\, - B_{g} \left(\kappa_{g}h^{\mu}{}_{\nu}-C^{2}\kappa_{f}\mathscr{h}^{\mu}{}_{\nu}\right) + \frac{\kappa_{g}}{2} {{\check{T}_{g}}{}}^{\mu}{}_{\nu}\, , \quad  A_{g}= -\frac{\Lambda_{g}}{\kappa_{g}}\, , \\
    \,^{(f)}\overline{\mathcal{E}}(\mathscr{h}^{\mu}{}_{\nu}) -\frac{\overline{R}_{f}}{4} \mathscr{h}^{\mu}{}_{\nu} = & \, B_{f} \left(\kappa_{g}h^{\mu}{}_{\nu}-C^{2}\kappa_{f}\mathscr{h}^{\mu}{}_{\nu}\right)+\frac{\kappa_{f}}{2\,C^{2}} {{\check{T}_{f}}{}}^{\mu}{}_{\nu}\, , \quad A_{f}=-\frac{\Lambda_{f}}{C^{2}\,\kappa_{f}},
\end{align}
where ${{\check{T}_{g}}{}}^{\mu}{}_{\nu}$ and ${{\check{T}_{f}}{}}^{\mu}{}_{\nu}$ are the trace-reversed energy-momentum tensors and
\begin{align}
    \nonumber \,^{(G)}\overline{\mathcal{E}}(h^{\mu}{}_{\nu}) \equiv  \frac{1}{2}\,^{(G)}\overline{\nabla}_{\lambda}\left[ \,^{(G)}\overline{\nabla}^{\mu}h^{\lambda}{}_{\nu}+\,^{(G)}\overline{\nabla}_{\nu}h^{\mu\lambda}-\,^{(G)}\overline{\nabla}^{\lambda}h^{\mu}{}_{\nu}\right] \, .
\end{align}

Equations in (\ref{eom_bg_spin2}) are interpreted as two spin-2 coupled equations, with massive and massless degrees of freedom living in a curved background. We can decouple (\ref{eom_bg_spin2}) to get to the mass basis. This defines the solutions of the gravity theory we are working on. By means of the Weyl double copy relation, in this paper we search for the gauge and scalar theories that are associated with certain solutions in bigravity. In Sec. \ref{Sec_WDC in BG}, we work with bigravity solutions and a Weyl double copy relation in this theory of massive gravity. 

Regarding the spinorial formalism in this theory, we apply the known definitions for the $g_{\mu\nu}$ in GR and extend them for the metric $f_{\mu\nu}$. Then, in bigravity we will consider two tetrads ${e_{g}}^{a}{}_{\mu}$ and ${{e}_{f}}^{a}{}_{\mu}$ aligned with the principal null directions and define the Infeld-van der Waerden symbols for both metrics, ${\sigma_{g}}^{\mu}_{A\dot{A}}$ and ${\sigma_{f}}^{\mu}_{A\dot{A}}$, so that we can convert tensors to spinors in both metrics. Then, we work with the Weyl spinors $\,^{(g)}{\Psi}_{ABCD}$ and $\,^{(f)}{\Psi}_{ABCD}$ and obtain the Weyl invariants in order to study how the Weyl spinor decomposes in the solutions we treat. The spinorial strengths are obtained by using $\,^{(0)}{\sigma_{g}}^{\mu}_{A\dot{A}}$ and $\,^{(0)}{\sigma_{f}}^{\mu}_{A\dot{A}}$ in (\ref{Fmunu_spinorial}). Finally, with these results we can give a Weyl double copy prescription for these solutions in bimetric massive gravity.

\subsection{Weyl double copy in BG}
For the solutions we studied, we obtained two cases which we can be classified according to the Petrov classification, depending on which the product of spinor basis $o_{A}, \iota_{B}$ appear on the Weyl spinor. Using the same notation, these ``type {\bf N}'' or type ``type {\bf D}'' solutions in bigravity are of the form: 
\begin{align}
    \label{Bigrav_classif}
    \nonumber &\text{Type {\bf N}:}\quad\,^{(G)}\Psi_{ABCD}=\left(\sum_{i}\psi_{G_{i}}+\sum_{j}\psi_{G_j}\right)o_{A}o_{B}o_{C}o_{D} \,,\quad \psi_{G_i}\equiv\,^{(G_i)}\psi_{4} \, , \\
    &\text{Type {\bf D}: }\quad\,^{(G)}\Psi_{ABCD}=\left(\sum_{i}6\psi_{G_{i}}+\sum_{j}\psi_{G_j}\right)o_{(A}o_{B}\iota_{C}\iota_{D)}\,,\quad \psi_{G_i}\equiv\,^{(G_i)}\psi_{2} \, , 
\end{align}
where 
\begin{align*}
    &\text{Type {\bf N}: } \ \ \ \ \ \psi_{G}=\sum_{i}\psi_{G_i}+\sum_{j}\psi_{G_j} \, , \quad \,^{(G)}\Psi_{ABCD}=\psi_{G} \, o_{A}o_{B}o_{C}o_{D}, \\
    &\text{Type {\bf D}: } \ \ \ \ \ \psi_{G}=\sum_{i}6\psi_{G_i}+\sum_{j}\psi_{G_j} \, ,\quad  \,^{(G)}\Psi_{ABCD}=\psi_{G}\,o_{(A}o_{B}\iota_{C}\iota_{D)}\,. 
\end{align*}
The index $i$ accounts for the modes that are present in GR: when $i=1$ we obtain the vacuum modes that we obtain in GR, and the term $i=2$ accounts for the source (matter) terms, i.e., the sourced Weyl double copy terms. The index $j$ accounts for the massive modes that appear in bigravity. Equation (\ref{Bigrav_classif}) is equivalent as saying that the Weyl spinor is decomposed in the term that appears in GR and the additional contribution of the massive modes 
\begin{align*}
    \,^{(G)}\Psi_{ABCD}= \,^{(G)}\Psi_{ABCD}^{\text{GR}}+\,^{(G)}\Psi_{ABCD}^{\text{Massive}},
\end{align*}
where $\,^{(G)}\Psi_{ABCD}^{\text{GR}}$ are the contributions from GR and $\,^{(G)}\Psi_{ABCD}^{\text{Massive}}$ are the additional terms in bigravity and account for the massive modes of the solution.

For the Weyl double copy of bigravity, Eq. (\ref{Bigrav_classif}) implies the following: we can construct the gauge and scalar theory solutions by considering the decompositions:
\begin{align}
    \,^{(G)}\Psi_{ABCD}^{\text{GR}}=\sum_{i=1}\frac{1}{S_{(G_i)}}f_{(AB}^{(G_i)}\cdot f_{CD)}^{(G_i)} \,, \quad \,^{(G)}\Psi_{ABCD}^{\text{\text{Massive}}}=\sum_{j}\frac{1}{S_{(G_j)}}f_{(AB}^{(G_j)}\cdot f_{CD)}^{(G_j)} \,.
\end{align}
For the solutions we present, we construct the gauge and scalar theories via:
\begin{align}
    \label{def_gauge_scalar_bg}
    \nonumber \text{Type {\bf N}}:  \ \ \ \ \quad f^{(G_i)}_{AB}&=\Theta_{G_i} o_{A}o_{B}\, , \quad S_{G_i}=\frac{(\Theta_{G_i})^{2}}{\psi_{G_{i}}}\,,\\
    \nonumber f^{(G_j)}_{AB}&=\Theta_{G_j} o_{A}o_{B}\, , \quad S_{G_i}=\frac{(\Theta_{G_j})^{2}}{\psi_{j}}\,.\\
    \text{Type {\bf D}}:  \ \ \ \ \ \quad f^{(G_i)}_{AB}&=\Theta_{G_i} o_{(A}l_{B)}\, , \quad S_{(G_i)}=\frac{(\Theta_{G_{i}})^{2}}{6\psi_{G_{i}}}\,,
\end{align}
similar to (\ref{def_gauge_scalar_th}). The prescription for type {\bf D} solutions is the same, as we have no massive contributions to the Weyl spinor. This is because the interaction term in these solutions is just the cosmological constant, which does not affect the invariants. In contrast, for type {\bf N} solutions, the interaction between the two metrics produces massive contributions to the Weyl spinor, so that massive fields arise in the Weyl double copy relations. Now we are ready to present the equations of motion. 

\subsection{WDC of Type N and Type D in BG}

\textbf{Type N solutions in bigravity}

The Weyl double copy of pp-waves and Siklos-(A)dS solution provides the scalars $S_{(G_{i})}$ and the spinor strengths $f^{(G_{i})}_{AB}$. In bigravity, we have additional fields coming from the massive contributions in the profile. The fields that appear in the recipe in (\ref{def_gauge_scalar_bg}) the massless fields $f^{(G_{1})}_{AB}$ and $S_{G_{1}}$
the matter fields $f^{(G_{2})}_{AB}$ and $S_{G_{2}}$ and the massive fields $f^{(G_j)}_{AB}$ and $S_{(G_{j})}$.  For the massless fields $f^{(G_{1})}_{AB}$ and $S_{(G_{1})}$ we have the following field equations:
\begin{align}
    \label{typeN_bg_1_eom}
    \nonumber \overline{\nabla}_{\mu}F_{(g_1)}^{\mu\nu}- \frac{\overline{R}}{6} A^{\nu}_{(g_1)}&=0\, , \quad &&\overline{\Box}\, S_{(g_1)}=0\,,\\
    \,^{(f)}\overline{\nabla}_{\mu}F_{(f_1)}^{\mu\nu}- \frac{\overline{\mathcal{R}}}{6} A^{\nu}_{(f_1)}&=0\, , \quad &&\,^{(f)}\overline{\Box}\, S_{(f_1)}=0\,.
\end{align}
For the matter fields $f^{(G_2)}_{AB}$ and $S_{(G_2)}$ the corresponding equations are
\begin{align}
    \label{typeN_bg_2_eom}
    \nonumber\overline{\nabla}_{\mu}F_{(g_2)}^{\mu\nu}- \left(\frac{\overline{R}}{6}+m^{2}_{\text{res}}\right) A^{\nu}_{(g_{2})}&=0\, , \qquad &&\left(\overline{\Box}+m^{2}_{\text{res}}\right)\, S_{(g_2)}=0\,, \\
    \,^{(f)}\overline{\nabla}_{\mu}F_{(f_2)}^{\mu\nu}-\left(\frac{\overline{\mathcal{R}}}{6}+\frac{1}{C^{2}}m^{2}_{\text{res}}\right)A^{\nu}_{(f_{2})}&=0\, , \qquad &&\left(\,^{(f)}\overline{\Box}+\frac{1}{C^{2}} m^{2}_{\text{res}}\right)\, S_{(f_2)}=0\,.
\end{align}
For the massive fields $f^{(G_j)}_{AB}$ and $S_{(G_j)}$ one has
\begin{align}
    \label{typeN_bg_pm_eom}
    \nonumber \overline{\nabla}_{\mu}F_{(g_j)}^{\mu\nu}- \left(\frac{\overline{R}}{6}+ \widehat{m}^{2}\right) A^{\nu}_{(g_j)} &=0\, , \qquad && \left(\overline{\Box}+\widehat{m}^{2}\right) S_{(g_j)}=0\,, \\
    \,^{(f)}\overline{\nabla}_{\mu}F_{(f_j)}^{\mu\nu}- \left(\frac{\overline{\mathcal{R}}}{6}+ \frac{\widehat{m}^{2}}{C^{2}}\right) A^{\nu}_{(f_j)} &=0\, , \qquad && \left(\,^{(f)}\overline{\Box}+\widehat{m}^{2}\right) S_{(f_j)}=0\,.
\end{align}
The field strengths $F^{(G_{I})}_{\mu\nu}$ are the tensor form of $f^{(G_I)}_{AB}$. Equivalently, the fields $A^{({G}_{I})}_{\mu}=\widetilde{\varphi}_{{G}_I}k^{({G})}_{\mu}$ with $I=\{i,j\}$ define the antisymmetric fields $F^{(G_{I})}_{\mu\nu}$ as usual. The fields are related through the set of equations  
\begin{align}
    \label{relations_bg_fields}
    \nonumber S_{(g_1)} &= \frac{{\kappa_{f}}^2}{{\kappa_{g}}^2} S_{(f_{1})}\,,\quad 
    &&f^{(g_1)}_{A B}=\frac{\kappa_{f}}{\kappa_{g}} f^{(f_{1})}_{AB}\,,\\
    \nonumber S_{(g_2)}&= \frac{\kappa_f}{\kappa_g}\frac{\Delta_g(m_\text{res})}{\Delta_f (m_\text{res})} S_{(f_2)}\,,\quad 
    &&f^{(g_2)}_{A B}=\frac{\Delta_{g}(m_\text{res})}{\Delta_f (m_\text{res})} \, f_{AB}^{(f_2)}\,,\\
    S_{(g_{j})} &= -C^2\,  S_{(f_j)}\, ,\quad
    &&f^{(g_{j})}_{A B}= - \frac{C^2 \kappa_g}{\kappa_f} f^{(f_{j})}_{AB}\,,
\end{align}
and the quantities $m_{\text{res}}$ and $\Delta_{G}(m_{\text{res}})$ depend on the matter content of the solution. 

We note that the massless and matter fields satisfy the same equations as in GR. The massive modes obey a Proca equation with a conformal curvature term for $A_{\mu}^{(G_{j})}$. The mass of the field  $A_{\mu}^{(g_{j})}$ is $\widehat{m}$ and  $A_{\mu}^{(f_{j})}$ have a mass of $\widehat{m}/C$. In the zeroth copy, and a massive Klein-Gordon equation for $S_{G_{j}}$, with identical mass $\widehat{m}$ in both cases. Then, for type {\bf N} solutions we have obtained massive modes using the Weyl double copy in bigravity. It is clear from (\ref{relations_bg_fields}) that there are only three independent modes, so that the equations of motion we obtain are also proportional.

\medskip

\noindent 
\textbf{Type D solutions in bigravity}

For type {\bf D} solutions we work with a dyonic Kerr-Newman-(A)dS and Pleba\'nski-Demia\'nski, both of which we have analyzed in the context of the WDC in GR. We also treat these solutions in bigravity and no massive modes appear, which means that for type {\bf D} solutions, $\psi_{G_j}=0$. The result we obtain after applying the Weyl double copy to these solutions is the same as in GR but for both metrics. Constructing the single and zeroth copy fields via (\ref{def_gauge_scalar_bg}), for the vacuum and matter fields we obtain
\begin{align}
\label{typeD_eom_1_bg} 
    \nonumber \,^{(G)}\overline{\nabla}_{\mu}F_{(G_1)}^{\mu\nu}&=0\, , \qquad &&\,^{(G)}\overline{\Box}\, S_{(G_1)}-\frac{\overline{R}_{G}}{6} S_{(G_1)}=0\,,\\
    \,^{(G)}\overline{\nabla}_{\mu}F_{(G_2)}^{\mu\nu}&=J^{\nu}_{(G_2)}\, , \qquad
    &&\,^{(G)}\overline{\Box}\, S_{(G_2)}-\frac{\overline{R}_{G}}{6} S_{(G_2)}=j_{(G_2)}\,,
\end{align}
where $J^{\nu}_{(G_2)}$ and $j_{(G_2)}$ are defined as in GR, for each one of the metrics.

As in GR, we work with solutions written in a KS form given by (\ref{ks_ansatz_bg}). This ansatz have been studied in the context of the double copy in bigravity in \cite{Garcia-Compean:2024zze,Garcia-Compean:2024uie}. For these solutions we coupled the matter in an independent way. If we use the effective metric to couple matter content, the result is that the charges/electromagnetic potentials of the metrics are proportional and not independent, as shown in \cite{Garcia-Compean:2024uie}. We proceed to review some solutions in bigravity and obtain their Weyl double copy.

\subsection{Examples}
Having studied some examples of the WDC in GR, we now present certain analogs of these solutions in bigravity. These solutions satisfy the formalism we have presented, where for type {\bf N} solutions we found massive modes entering the WDC prescription, in contrast to the case of type {\bf D} solutions, where no massive contributions are present. 

\subsubsection{Bigravitational pp-waves}
We consider pp-waves solutions in bigravity as in \cite{Ayon-Beato:2018hxz}. The background $d\overline{s}^2$ and the null vector $k_{\mu}$ are the same as in the GR solution. The tetrad for $g_{\mu\nu}$ is the same as in (\ref{tetrad_pp}), and for $f_{\mu\nu}$ thus we have
\begin{align}
\label{tetrad_pp_bg}
    \left.\mathscr{e}^a{}_{\mu}=\left(
\begin{array}
{cccc}\frac{1}{2}(1-C^{2}\kappa_f\phi_{f}) & 1 & 0 & 0 \\
\frac{1}{2}(1-C^{2}\kappa_f\phi_{f}) & -C^2 & 0 & 0 \\
0 & 0 & C & 0 \\
0 & 0 & 0 & C
\end{array}\right.\right).
\end{align}
The profiles as in \cite{Ayon-Beato:2018hxz} can be written as:
\begin{align}
   \nonumber  \phi_{g_1}&=\frac{\kappa^2}{\kappa_{g}(C^{2}{\kappa_{g}}^{2}+{\kappa_{f}}^{2})}\,f_{2}(u)(x^{2}-y^{2})=\frac{\kappa_{f}}{\kappa_{g}}{\phi}_{f_1}\,,\\
    \phi_{g_{j}}&=-\frac{2C^2 \kappa_{g}}{C^2 {\kappa_{g}}^2+{\kappa_{f}}^2} \, H_{j}(u) \, e^{\widehat{m}x_{j}}=-\frac{C^2\kappa_{g}}{\kappa_{f}}\phi_{f_j},
\end{align}
where  $x_3=x=-x_5$ and $x_4=y=-x_6$. We used the arbitrary functions $\vec{H}_{+}(u)=(H_{3}(u),H_{4}(u)), \vec{H}_{-}(u)=(H_{5}(u),H_{6}(u))$ and $e^{\pm\widehat{m}\vec{x}}=(e^{\pm\widehat{m}x},e^{\pm\widehat{m}y})$. Then, the components for the massive modes that we use in this solution correspond to the values of $j=3,4,5,6$.

For a type {\bf N} solution, we use the Weyl invariant $\,^{(G)}\psi_4=\psi_{G_1}+\psi_{G_{2}}$, and because we are in vacuum, $\psi_{G_{2}}=0$. For each metric, the Weyl invariant and the Weyl spinor can expressed as
\begin{align}
    \psi_{G}=\frac{\kappa_{G}}{2} \left(\partial_{x} + i \,  \partial_{y}\right)^{2}\phi_{G}(u,x,y)\, , \ \ \ \quad \,^{(G)}\Psi_{ABCD}=\psi_{G} o_{A}o_{B}o_{C}o_{D}\,,
\end{align}
so that $\psi_{G}$ is decomposed in the following form 
{\small
\begin{align}
    \nonumber \psi_{G_{I}}=\frac{\kappa_{G}}{2} \left(\partial_{x} + i \,  \partial_{y}\right)^{2}\phi_{G_1}(u,x,y)\, ,
\end{align}
}where $I$ stands for GR and the massive modes. Then we have
\begin{align*}
    \psi_{g_1}=\frac{2 \kappa^{2}}{C^2 {\kappa_{g}}^2+{\kappa_{f}}^2}f_{2}(u) = \psi_{f_1}\,,\qquad \ \ \ 
    \psi_{g_{j}}=\eta_{j} \frac{\widehat{m}^2}{2}\kappa_{g} \phi_{g_{j}}=\frac{C^2 {\kappa_{g}}^2}{{\kappa_{f}}^2} \,\psi_{f_{j}} \, ,
\end{align*}
where $\eta_{3}=-\eta_{4}=1=\eta_{5}=-\eta_{6}$. 

Now we construct the scalar functions $\Theta_{G_I}$ and the strength spinors $f^{(G_I)}_{AB}$ in bigravity. We can map  each of the terms in $\phi_{G}$ to a function $\widetilde{\varphi}_{G_{I}}$, introducing the arbitrary functions of the retarded time $u$}, $\widetilde{\mathrm{f}}_{2}(u)$ and $\vec{\mathrm{h}}_{+}(u)$, where  $\vec{{\mathrm{h}}}_{\pm}(u)=\big(\widetilde{\mathrm{h}}_{3}(u),\widetilde{\mathrm{h}}_{4}(u)\big)$ and $\vec{{\mathrm{h}}}_{-}(u)=\big(\widetilde{\mathrm{h}}_{5}(u),\widetilde{\mathrm{h}}_{6}(u)\big)$. Then, with these functions we define $\Theta_{G_{I}}$ that appear in the strength spinor as
\begin{align}
    \nonumber \widetilde{\varphi}_{g_1}&=\frac{\kappa^2}{\kappa_{g}(C^{2}{\kappa_{g}}^{2}+{\kappa_{f}}^{2})}\,\widetilde{\mathrm{f}}_{2}(u)(x^{2}-y^{2})=\frac{\kappa_{f}}{\kappa_{g}}{}\widetilde{\varphi}_{f_1}, \,\\
    \nonumber \widetilde{\varphi}_{g_{j}}&=-\frac{2 C^2 \kappa_g}{C^{2}{\kappa_{g}}^{2}+{\kappa_{f}}^{2}}\,{\widetilde{\mathrm{h}}}_{j}(u) e^{\widehat{m}{x_j}} =-\frac{C^{2}\kappa_{g}}{\kappa_{f}}\widetilde{\varphi}_{f_{j}}\, \qquad \Theta_{G_I}=-i\left[\left(\partial_{x} + i \,  \partial_{y}\right)\widetilde{\varphi}_{G_I} \right]\, .
\end{align}
Once this is defined, we can use the Weyl spinor and construct the scalar and gauge fields. For the metric $g$, the massless term is mapped to
{\small
\begin{align}
    \nonumber &S_{(g_1)} =\,\frac{(\Theta_{g_{1}})^{2}}{\psi_{g_1}}=- \frac{2{\kappa}^{2}}{{\kappa_{g}}^2\,(C^{2}\kappa_g^{2}+\kappa_f^{2})}\frac{\widetilde{\mathrm{f}}_{2}(u)^{2} (x-i y)^2}{f_2(u)} =- \frac{2}{\kappa_g}\frac{\widetilde{\mathrm{f}}_{2}(u)\, (x-i y)^2}{f_{2}(u)\,(x^2-y^2)} \widetilde{\varphi}_{g_1}\, , \\
    \nonumber &f^{(g_1)}_{A B}= \Theta_{g_{1}} o_{A}o_{B} = \left(i\,\frac{{2\kappa}^{2}}{\kappa_g\, ({C^{2}\kappa_g^{2}+\kappa_f^{2}}^{2})}\widetilde{\mathrm{f}}_{2}(u)\,(x-i y)\right)o_{A}o_{B}=\left(-2 i \,\left(\frac{x-iy}{x^2-y^2}\right)\,\widetilde{\varphi}_{g_1}\right) \, o_A o_B\,,
\end{align}}
which is a similar result as in Einstein's theory. For the massive modes we get
{\small\begin{align} 
    \nonumber &S_{(g_{j})} =\,\frac{(\Theta_{g_{j}})^{2}}{\psi_{g_{j}}} = \,\frac{2C^{2}}{C^{2}\kappa_g^{2}+\kappa_f^{2}}\frac{{\widetilde{\mathrm{h}}}_{j}(u)^{2} \, e^{\widehat{m}x_j}}{{H}_{j}(u)}=-\frac{2}{\kappa_{g}}\frac{\widetilde{\mathrm{h}}_j(u)}{h_j (u)}\widetilde{\varphi}_{g_j}\, , \\
    \nonumber &f^{(g_{j})}_{A B}=\Theta_{G_{j}}o_{A}o_{B}=\left(- \,i^{j}\,\frac{C^{2}{\kappa}_{g}}{C^{2}\kappa_g^{2}+\kappa_f^{2}}\,\widehat{m}\,\widetilde{\mathrm{h}}_{j}\,e^{\widehat{m}x_j}\right)o_{A}o_{B}=\left( i^j\, \widehat{m} \widetilde{\varphi}_{g_j} \right) o_A o_B \, .
\end{align}}These results are related to the ones of the metric $f$ by (\ref{relations_bg_fields}).

For pp-waves, we have vanishing cosmological constants for the two metrics, $\Lambda_{G}=0$. To write down the equations of motion for the single and zeroth copies, we will use the null vectors in the background Minkowski space. Thus we define
\begin{align}
    \nonumber\overline{k}^{(g)}_{\mu}dx^{\mu}=-du \, , \quad \ \ \ \ \ \overline{k}^{(f)}_{\mu}dx^{\mu}=-C du,\, 
\end{align} 
where $k_{\mu}=\overline{k}^{(g)}_\mu$. Then, we can use (\ref{Fmunu_spinorial}) with $\,^{(0)}{\sigma_{G}}^{\mu}_{A\dot{A}}$ and obtain the field $F^{(G_{i})}_{\mu\nu}$ equivalently defined via the potential $A^{({G}_I)}_{\mu}=\widetilde{\varphi}_{G_I}\overline{k}^{({G_{0}})}_{\mu}$ where $G_{0}$ stands for the flat metric of the solution. Then, with $\Theta_{i}$ and $\psi_{G_i}$ we obtain $S_{(G_{i})}$.

The equations of motion we obtain for massless $f^{(G_{1})}_{AB}, S_{(G_{1})}$ and massive fields $f^{(G_{j})}_{AB}, S_{(G_{j})}$ for both metrics are equations (\ref{typeN_bg_pm_eom}) with $\Lambda_{G}=0$, i.e., the same equations but in a Minkowski background. Then, there is no scalar curvature term and the differential operator is the one associated to the flat background, which results in
\begin{align}
    \nonumber \,^{(G_0)}\nabla_{\mu}F_{(G_1)}^{\mu\nu}=0\, , \qquad \,^{(G_0)}\Box\, S_{(G_1)}=0, \,
\end{align}
for the massless modes. Moreover, the equations for the massive modes are given by 
\begin{align}
    \nonumber \,^{(g_{0})}{\nabla}_{\mu}F_{(g_j)}^{\mu\nu}- \widehat{m}^{2}\, A^{\nu}_{(g_j)} &=0\, , \quad && \left(\,^{(g_{0})}{\Box}+\widehat{m}^{2}\right) S_{(g_j)}=0\,, \\
    \,^{(f_0)}{\nabla}_{\mu}F_{(f_j)}^{\mu\nu}- \frac{\widehat{m}^{2}}{C^{2}}\,A^{\nu}_{(f_\pm)} &=0\, , \quad && \left(\,^{(f_0)}{\Box}+\widehat{m}^{2}\right) S_{(f_j)}=0, \,
\end{align}
where the subscript ``$(G_0)$'' denotes that the operator is constructed using the the corresponding flat metric for a given spacetime. In this situation the relation between the Weyl and the Kerr-Schild double copy is similar to the ones in GR.
\subsubsection{Bigravitational AdS waves}
We turn our attention to AdS waves in massive (bi-)gravity. Bi-gravitational waves propagating in a flat space were studied in \cite{Mohseni:2012ug} and the case of AdS waves in bigravity was presented in \cite{Ayon-Beato:2018hxz}. We shall analyze the latter solution using the Weyl double copy.

We use the same background and null vector as in GR. The tetrad for $g_{\mu\nu}$ is the same as in (\ref{tetrad_Siklos}), and for $f_{\mu\nu}$ we have
\begin{align}
\label{tetrad_siklos_bg}
    \left.\mathscr{e}^a{}_{\mu}=\frac{l}{y}\left(
\begin{array}
{cccc}\frac{1}{2}(1+C^{2}\kappa_f \phi_{f}) & 1 & 0 & 0 \\
\frac{1}{2}(1-C^{2}\kappa_f\phi_{f}) & -C^2 & 0 & 0 \\
0 & 0 & C & 0 \\
0 & 0 & 0 & C
\end{array}\right.\right)\, .
\end{align}
In \cite{Ayon-Beato:2018hxz}, massive and massless profiles were presented for the vacuum and non-vacuum cases. For the massive profiles, the scalar functions $\phi_{G}$ in the metric that describe bigravitational waves in vacuum propagating in an AdS background are defined as
{\small
\begin{align*}
    \phi_{g_1}&=\frac{\kappa^{2}}{\kappa_g({\kappa_{f}}^{2}+C^{2}{\kappa_{g}}^{2})} f_{3}(u)\left(\frac{y}{l}\right)^{3}=\frac{\kappa_g}{\kappa_f}\,\phi_{f_1}\,,\\
    \phi_{g_\pm}&=-\frac{C^{2} \kappa_g}{{\kappa_{f}}^{2}+C^{2}{\kappa_{g}}^{2}} h_{\pm}(u)\left(\frac{y}{l}\right)^{\rho_{\pm}}\ =- \frac{C^2\kappa_g}{\kappa_f}\,\phi_{f_\pm}\,,
\end{align*}}where we have used the arbitrary functions $f_{3}(u)$ and $h_{\pm}(u)$. The indices for the massive modes that we employ are $j=\{+,-$\}. 

In vacuum one has that $\psi_{G_{2}}=0$, so that in this case the Weyl invariant is $\,^{(G)}\psi_4=\psi_{G_1}$ and $\psi_{G_{2}}=0$. The Weyl invariants and spinor are given by
\begin{align}
    \psi_{G}=\frac{\kappa_G}{2} \left(\frac{y}{l}\right)^{2}\left(\left[\left(\partial_{x} + i \,  \partial_{y}\right)^{2}\right]\phi_{G}(u,x,y)\right)\, , \quad \,^{(G)}\Psi_{ABCD}=\psi_{G} o_{A}o_{B}o_{C}o_{D}\,,
\end{align}
and the Weyl invariants can be written as
{
\begin{align}
    \nonumber \psi_{G_{1}}=\frac{\kappa_G}{2} \left(\frac{y}{l}\right)^{2}\left[\left(\partial_{x} + i \,  \partial_{y}\right)^{2}\right]\phi_{G_1}\, ,\quad
    \psi_{G\pm}=-\frac{\kappa_G}{2}y^2\frac{\left(\widehat{m}^{2}+ \frac{2}{l^{2}}\rho_{\pm}\right)}{(1-\rho_{\pm})\rho_{\pm}}\left[\left(\partial_{x} + i \,  \partial_{y}\right)^{2}\right]\phi_{G_\pm}\, .
\end{align}
These expressions allows us to write down
\begin{align*}
    \psi_{g_1}=\Lambda\, \kappa_{g} \phi_{g_1}=\psi_{f_1}\, ,\quad 
    \psi_{g_\pm}=-\frac{\kappa_g}{2}\left(\widehat{m}^{2}+ \frac{2}{l^{2}}\rho_{\pm}\right) \, \phi_{g_\pm} =-\frac{C^2 {\kappa_{g}}^2}{{\kappa_{f}}^2}\psi_{f_\pm}.
\end{align*}
We associate $\phi_{G_i}$ to the functions $\widetilde{\varphi}_{G_{i}}$ using the arbitrary functions of the retarded time $u$, $\widetilde{\mathrm{f}}_{3}(u)$ and $\widetilde{\mathrm{h}}_{\pm}(u)$, which defines $\Theta_{G_I}$ as
\begin{align}
    \nonumber \widetilde{\varphi}_{g_1}&=\left. \phi_{g_1} \right|_{f_{3}\rightarrow \widetilde{f}_{3}}=\frac{\kappa_g}{\kappa_f}\widetilde{\varphi}_{f_1}\,,\quad &&\Theta_{G_1}=i\frac{2}{3}\frac{y}{l}\left[\left(\partial_{x} + i \,  \partial_{y}\right)\widetilde{\varphi}_{G_1} \right]\, ,\\
    \widetilde{\varphi}_{g_\pm}&=\left. \phi_{g_1} \right|_{\mathrm{h}_{\pm}\rightarrow \widetilde{\mathrm{h}}_{\pm}} =- \frac{C^2\kappa_g}{\kappa_f}\,\widetilde{\varphi}_{f_\pm},\, \quad &&\Theta_{G_\pm}=-i\frac{(1-\rho_{\pm})}{\rho_{\pm}}\frac{y}{l}\left[\left(\partial_{x} + i \,  \partial_{y} \right)\widetilde{\varphi}_{G_\pm} \right].\,
\end{align}
With these results, for the metric $g$ we have the massless terms
{\small
\begin{align}
    \nonumber S_{(g_1)} =-\frac{4}{3 \, \kappa_g}\frac{\widetilde{\mathrm{f}}_{3}(u)}{f_3(u)} \, \widetilde{\varphi}_{g_1}\, , \qquad 
    f^{(g_1)}_{A B} \left(-\frac{2}{l} \widetilde{\varphi}_{g_1}\right)\, o_{A}o_{B}\,,
\end{align}}
as in GR. The massive contributions give rise to the fields
{\small\begin{align} 
    \nonumber S_{(g_\pm)} =\frac{2}{\kappa_g}\left(\frac{1-\rho_\pm}{\rho_\pm}\right)\frac{\widetilde{\mathrm{h}}_\pm(u)}{h_\pm (u)} \,  \widetilde{\varphi}_{g_\pm}
    \, , \ \ \ \ \ \qquad
    f^{(g_\pm)}_{A B}=\left( \frac{(1-\rho_\pm)}{l}\, \widetilde{\varphi}_{g_\pm}\right)o_{A}o_{B} \, .
\end{align}}
These expressions are related to the fields in $f$ as in (\ref{relations_bg_fields}).
}

By implementing (\ref{Fmunu_spinorial}), we recover $F^{(G_{I})}_{\mu\nu}$ defined by $A^{({G}_I)}_{\mu}=\widetilde{\varphi}_{I}\overline{k}^{({G})}_{\mu}$, where in this case we use the null vectors of AdS space \begin{align}
    \nonumber \overline{k}^{(g)}_{\mu}dx^{\mu}=\frac{l}{y}du \, , \ \ \ \ \ \ \ \quad \overline{k}^{(f)}_{\mu}dx^{\mu}=C \frac{l}{y} du.\, 
\end{align} 
Then, these fields have the same behavior as in GR for both metrics, as written in (\ref{typeN_bg_1_eom}) with an AdS background.

In the massless limit, when $\widehat{m}=0$, the interaction between the metrics is just the cosmological constant. The $\phi_{G_-}$ mode vanishes and the $\phi_{G_+}$ mode has the same form as $\phi_{G_1}$. We end with only one massless mode ${\phi}_{G_1}'=\phi_{G_1}+\phi_{G_+}$ which leads to Maxwell and Klein-Gordon equations in an AdS background.

We note the relation between the Kerr-Schild and Weyl double copy
\begin{align*}
    \left. \mathfrak{Re}(S_{(G_1)})\right|_{\widetilde{\mathrm{f}}_{2} \rightarrow f_2}=-\frac{4}{3\,\kappa_G} \,\phi_{G_1}\,,\quad\left. \mathfrak{Re}(S_{(G_j)})\right|_{\widetilde{\mathrm{h}}_{j} \rightarrow h_j}=-\frac{2}{\kappa_G} \frac{(1-\rho_\pm)^2}{l^2(\widehat{m}^2+
    \frac{2}{l^2}\,\rho_\pm)}\,\phi_{G_j}\,.
\end{align*}
Even in the case of pp-waves, we could have defined the scalar fields as a sum of the massive modes from the start. The resulting equations of motion for $S_{(G_j)}$ deviates from the massive Klein-Gordon equation by an additional term. Then, we defined each of the terms in the profile as a field in the Weyl double copy prescription. We note that the Kerr-Schild form provides a natural way of separating each of the contributions in the Weyl spinor \cite{Caceres:2025eky}, which permit us to construct the Weyl double copy. 

\medskip

\subsubsection{Bigravitational Siklos-Maxwell waves}
We turn our attention to non-vacuum solutions. We can couple Siklos bigravitational waves to the the electromagnetic field using the effective metric following \cite{Ayon-Beato:2018hxz}.
We define
\begin{align}
    \nonumber \Delta_{g}(m_\text{er})&\equiv\frac{\kappa_g}{4\pi}\frac{\left[\widehat{m}^{2}(\alpha+C\beta)\,\kappa_f^{2}-m^{2}_{\text{er}}\,\alpha\left(C^{2}\kappa_g^{2}+\kappa_f^{2}\right)\right]}{l^{2} (\alpha+C\beta) \left(C^{2}\kappa_g^{2}+\kappa_f^{2}\right)\left(\widehat{m}^{2}- m_{\text{er}}^{2}\right)}\, , \\
    \Delta_{f}(m_\text{er})&\equiv\frac{\kappa_f}{4\pi}\frac{\left[\widehat{m}^{2}(\alpha+C\beta)\,C^{2}\kappa_g^{2}-m^{2}_{\text{er}}\,C\beta\left(C^{2}\kappa_g^{2}+\kappa_f^{2}\right)\right]}{ l^{2} C^{2}(\alpha+C\beta)\left(C^{2}\kappa_g^{2}+\kappa_f^{2}\right)\left(\widehat{m}^{2}-m_{\text{er}}^{2}\right)}\,,
\end{align}
which allow us to write down the profiles of the metrics as
{\small
\begin{align}
    \nonumber 
    \phi_{G_2}&=
    \Delta_{G}(m_\text{er})\,D_{2}(u)^{2}y^{4}\,,\qquad m^{2}_{\text{er}}=\frac{4}{l^2}=\frac{2}{3}\left(\rho_{(g_2)}^{\text{grav}}- \Lambda_g\right)=\frac{2\,C^2}{3}\left(\rho_{(f_2)}^{\text{grav}}- \Lambda_f\right)\,,
\end{align}}
where $\phi_G=\phi_{G_{1}}+\phi_{G_2}+\sum\phi_{G_j}$ and $m_\text{er}$ is defined as before. We can also have the case for electromagnetic resonance for $m_\text{res}=m_{\text{er}}$,
which changes the behavior of the profiles we are considering. 
The Weyl invariants and spinor has the same form
\begin{align}
    \psi_{G}=\frac{\kappa_G}{2} \left(\frac{y}{l}\right)^{2}\left[\left(\partial_{x} + i \,  \partial_{y}\right)^{2}\right]\phi_{G}(u,x,y) \, , \quad \,^{(G)}\Psi_{ABCD}=\psi_{G} o_{A}o_{B}o_{C}o_{D}\,,
\end{align}
and we are now accounting for the matter content by means of $\phi_{G_2}$. Then 
\begin{align}
    \psi_{G_{2}}=\frac{\kappa_G}{2} \left(\frac{y}{l}\right)^{2}\left[\left(\partial_{x} + i \,  \partial_{y}\right)^{2}\right]\phi_{G_2} \,= 2 \Lambda \,\kappa_{G}\,\phi_{G}.
\end{align} 
We map the matter part $\phi_{G_2}$ to the functions $\widetilde{\varphi}_{G_2}$, which allow us to write $\Theta_{G_2}$ that appears in $f^{(G_2)}_{AB}$. Then, we have that
\begin{align*}
    \widetilde{\varphi}_{G_2}(u,x,y)= \Delta_{G}(m_\text{er})\,\widetilde{D}_{2}(u)^{2}y^{4}\,,\qquad \Theta_{G_2}=\frac{3i}{4}\left(\frac{y}{l}\right)\left[\left(\partial_{x} + i \,  \partial_{y} \right)\widetilde{\varphi}_{G_2} \right]. \, 
\end{align*}
For the matter part of the metrics $g$ and $f$, we obtain the following fields
{\small
\begin{align}
    \nonumber S_{(g_2)}=-\frac{3}{2\,\kappa_g} \frac{\widetilde{D}_{2}(u)^{2}}{D_2(u)} \phi_{g_2} \,,\quad \ \ \ \ \ 
    f^{(g_2)}_{A B}= \left(-\frac{3}{l}\, \phi_{g_i} \right)o_{A}o_{B},\,
\end{align}}which are related to the fields for $f$ as in (\ref{relations_bg_fields}). The equations of motion for $G=g,f$ that we obtain are equations (\ref{typeN_bg_2_eom}) with $m_{\text{res}}=m_{\text{er}}$. $S_{G_2}$ and $\phi_{G_2}$ are related precisely as in GR.

\medskip
\subsubsection{Bigravitational Siklos-Scalar waves}

We now consider coupling a scalar field $\Phi$ to the solution in bigravity using the effective metric. First of all we define
\begin{align}
    \nonumber \Delta_{g}(m_\text{sr})&\equiv-\frac{1}{2}\frac{\left[\widehat{m}^{2}(\alpha+C\beta)\,\kappa_f^{2}-m^{2}_{\text{sr}}\,\alpha\left(C^{2}\kappa_g^{2}+\kappa_f^{2}\right)\right](\alpha+C\beta)\,\kappa_g}{\left(C^{2}\kappa_g^{2}+\kappa_f^{2}\right)\left(\widehat{m}^{2}- m_{\text{sr}}^{2}\right)}\, , \\
    \Delta_{f}(m_\text{sr})&\equiv-\frac{1}{2}\frac{\left[\widehat{m}^{2}(\alpha+C\beta)\,C^{2}\kappa_g^{2}-m^{2}_{\text{sr}}\,C\beta\left(C^{2}\kappa_g^{2}+\kappa_f^{2}\right)\right] (\alpha+C\beta)\,\kappa_f}{C^{2}\left(C^{2}\kappa_g^{2}+\kappa_f^{2}\right)\left(\widehat{m}^{2}-m_{\text{sr}}^{2}\right)}\,,
\end{align}
so that the matter profiles are given by \cite{Ayon-Beato:2018hxz}
{\small
\begin{align}
    \nonumber \phi_{G_2}&=
    \Delta_{G}(m_\text{sr})\,\dot{\Phi}(u)^{2}y^{2}\,,\qquad m^{2}_{\text{sr}}=-\frac{2}{l^2}=\frac{1}{3}\left(\rho_{(g_2)}^{\text{grav}}- \Lambda_g\right)=-\frac{C^2}{3}\left(\rho_{(f_2)}^{\text{grav}}- \Lambda_f\right)\,.
\end{align}}For the case of scalar resonance, where $\widehat{m}=m_{sr}$, the profiles are modified to a logarithmic behavior. We consider the massive profiles for arbitrary mass $\widehat{m}$.

The Weyl invariant corresponding to the matter contribution is given by
\begin{align*}
\psi_{G_2}= \frac{1}{3} \Lambda \,\kappa_{G} \phi_{G}\,.
\end{align*}
We also define the functions $\widetilde{\varphi}_{G_2}$ associated to the profiles $\phi_{G_2}$ and write $\Theta_{G_2}$ as
\begin{align*}
    \widetilde{\varphi}_{G_2}(u,x,y)&= \Delta_{G}(m_\text{sr})\,\dot{\widetilde{\Phi}}(u)^{2}y^{2}\,, \qquad \Theta_{G_2}=\frac{i}{2}\left(\frac{y}{l}\right)\left[\left(\partial_{x} + i \,  \partial_{y} \right)\widetilde{\varphi}_{G_2} \right] \, , 
\end{align*}
which allows to write the matter strength spinor $f^{(G_2)}_{AB}$. We have used the arbitrary function $\widetilde{\Phi}(u)$. The vacuum and massive modes are also identified with their respective functions $\widetilde{\varphi}_{G_I}$ as before.

For the matter part of the Weyl double copy we obtain
\begin{align}
    \nonumber S_{(G_2)}=-\frac{1}{\kappa_{G}}\frac{\dot{\widetilde{\Phi}}(u)}{\dot{\Phi}(u)}\widetilde{\varphi}_{G_2}
    \,,\qquad
    f^{(G_2)}_{A B}= \left(-\frac{1}{l}\widetilde{\varphi}_{G_2}\right)o_{A}o_{B}\,.
\end{align}
The fields also relate to the ones in $f$ via (\ref{relations_bg_fields}). Then, for $G=g,f$, the equations of motion are (\ref{typeN_bg_2_eom}) with $m_{\text{res}}=m_{\text{sr}}$. $S_{G_2}$ and $\phi_{G_2}$ are also related as in GR.

\subsubsection{Dyonic Kerr-Newman-(A)dS in BG}

In bigravity we can have configurations of two rotating black holes with mass $m_{g}$ and $m_{f}$ and equal rotation parameter $a$ in asymptotically flat space \cite{Babichev:2014tfa}. The asymptotically (A)dS case was first found in \cite{Ayon-Beato:2015qtt}.

Kerr-Newman-(A)dS solution in bigravity \cite{Ayon-Beato:2015qtt} written in its Kerr-Schild form (\ref{ks_ansatz_bg}) uses the same background metric and null vector as in GR. We can generalize the solution in \cite{Ayon-Beato:2025ahb} given the Kerr-Schild form, so that the dyonic KN-(A)dS solution in bigravity, with matter independently coupled has the following profiles and electromagnetic potentials
\begin{align}
    \nonumber \phi_{G} =-\frac{\kappa_{G}}{2}\frac{[2m_G r-(e_G ^2+p_G^2)]}{8\,\pi\,\Sigma} \, , \quad {A_f}_{\mu}\equiv \frac{C e_{G}\,r}{\Sigma}k_{\mu}- \frac{C p_G \cos\theta}{\Sigma} k^*_\mu\,,
\end{align}
and ${A_g}_{\mu}$ is defined as in GR. The tetrad for the metric $g$ is the same as in (\ref{tetrad_dKNg}), and for the metric $f$ the solution is given by
{\begin{align}
\nonumber
    \mathscr{e}^a{}_{\mu}=C \,
\begin{pmatrix}
\dfrac{\Delta}{\Omega}\sqrt{\dfrac{\Delta_{f_r}}{\Sigma}} & \sqrt{\dfrac{\Sigma}{\Delta_{f_r}}}\dfrac{\Sigma\,\kappa_{f}\, \phi_f }{\left(r^2+a^2\right)\,\xi} & 0 & -\sqrt{\dfrac{\Delta_{f_r}}{\Sigma}} \dfrac{a \sin^2 \theta}{\Omega}  \\
0 & \dfrac{\Sigma}{\Delta_{f_r}} & 0 & 0 \\
0 & 0 & \sqrt{\dfrac{\Sigma}{\Delta}} & 0 \\
-\sqrt{\dfrac{\Delta}{\Sigma}} \dfrac{\xi a \sin \theta}{\Omega} \, & 0 & 0 & \sqrt{\dfrac{\Delta}{\Sigma}} \dfrac{\sin \theta \, \left(r^{2}+a^2\right)}{\Omega}
\end{pmatrix}\, ,
\end{align}}
with $\Delta_{f_r}=(r^2+a^2)\,\xi+\kappa_{f}\phi_f \,\Sigma$. 
Then, the gauge and scalar fields are written as
\begin{align}
    \nonumber S_{(f_1)} &=\frac{1}{6}\frac{16 \pi}{\kappa_{f}^{2}}\frac{\widetilde{m}^{2}_f}{C^2\,m_f\,\left(r-i a \cos\theta\right)} \, , \quad
    f^{(f_1)}_{A B}= -\frac{\widetilde{m}_f}{C^{2}\,\left(r-i a \cos\theta\right)^{2}}\,o_{(A}l_{B)} \, , \\
    \nonumber S_{(f_2)} &=-\frac{1}{6}\frac{16 \pi\,C^2}{\kappa_{f}^{2}}\frac{(\widetilde{e}_{f}^2+\widetilde{p}_{f}^2)^{2}}{(e^{2}_f+p^{2}_f)\,\left(r-ia\cos\theta\right)\left(r+ia\cos\theta\right)}, \, \\ 
    f^{(f_2)}_{A B} &= \,\frac{(\widetilde{e}_{f}^2+\widetilde{p}_{f}^2)}{\left(r-ia\cos\theta\right)^{2}\left(r+ia\cos\theta\right)}\,o_{(A}l_{B)},
\end{align}
with $A^{(g_1)}_{\mu}$ as in GR and a similar expression for $A^{(f_1)}_{\mu}$ with $\widetilde{m}_f$. Moreover, the sources can be expressed as
{\begin{align}
    \nonumber \,^{(0)}{J}^{\nu}_{(f_2)}&=\frac{\widetilde{e}_{f}^2+\widetilde{p}_{f}^2}{C^2\,\left(r^2+a^2\cos^2\theta\right)^{3}}\left[(r^2-a^2\cos^2\theta+2a^2 )\,\delta^{\nu}{}_{t}+2 a\,\delta^{\nu}{}_{\phi}\right],\, \\ 
    \nonumber \widetilde{\Lambda}_{(f_2)}^{\nu}&= -\frac{1}{3}\frac{C^2\,(r^2-a^2 \cos^2\theta)}{(r^2+a^2\cos^2\theta)^3} \,\Lambda\, a \, (\widetilde{e}_{f}^2+\widetilde{p}_{f}^2)\,\delta^{\nu}_{\phi},\,\\
    \nonumber \,^{(0)}j_{(f_2)}&=\frac{4}{3}\frac{64\pi^2\,C^{2}}{\kappa_{f}^4}\frac{(\widetilde{e}_{f}^2+\widetilde{p}_{f}^2)^2}{(e^{2}_f+p^{2}_f)^2}\,\,^{(0)}\rho_{(f_2)}^{\text{grav}},  \,\\
    \widetilde{\lambda}_{(f_2)}&=-
    \frac{4}{3}\frac{64\pi^2 \,C^{4}}{\kappa_{f}^4}\frac{(\widetilde{e}_{f}^2+\widetilde{p}_{f}^2)^2}{(e^{2}_f+p^{2}_f)^2}\, \,^{(0)}\rho^{\text{grav}}_{(f_2)} \left(\frac{1}{3}\frac{(r^2-a^2\cos^2{\theta}- 2\,r^2\cos^2\theta)}{r^2-a^2\cos^2\theta +2 a^2}\,\Lambda\, a^2\right).\,
\end{align}}
These fields have the same behaviour as in GR, and follow the equations (\ref{typeD_eom_1_bg}).

\subsubsection{Pleba\'nski-Demia\'nski in BG}
In this case, we use the same background metric $\overline{g}_{\mu\nu}$ and null vectors. The scalar profiles for both metrics are given by
\begin{align}
    \label{profiles_PD_bg}
    \phi_{G}= -\frac{\kappa_{G}}{2}\frac{2m_{G}p-{e_{G}}^2}{8\,\pi\,(q^2-p^2)}\, , \quad \chi_{G}=-\frac{\kappa_{G}}{2}\frac{2n_{G}q+{\mathrm{g}_{G}}^{2}}{8\,\pi\,(q^2-p^2)}\,, \quad  {A_{f}}_{\mu}=\frac{C}{q^2-p^2}\left(e_{f}\, q \,l_{\mu}+\mathrm{g}_{f}\,p\,k_{\mu}\right), \,
\end{align}
where $m_{g}=m, n_{g}=m$ and so on. We consider independent matter couplings to the metrics: the electromagnetic potential for $g_{\mu\nu}$ is the same in GR, ${A_{g}}_{\mu}$, and the potential for $f_{\mu\nu}$ is $ {A_{f}}_{\mu}$. The tetrad for $g_{\mu\nu}$ is the same as in (\ref{tetrad_PDg}), and for $f_{\mu\nu}$ we have
\begin{align}
\label{tetrad_PD_bg_f}
    \mathscr{e}^a{}_{\mu}=\frac{C}{\sqrt{2}}
\begin{pmatrix}
1+\Delta_{q_f} & p^2(1+\Delta_{p_f}) & 0 & 1 \\
1-\Delta_{q_f} & p^2(1-\Delta_{q_f}) & 0 & -1 \\
-(1+\Delta_{p_f}) & -q^2(1+\Delta_{p_f}) & 1 & 0 \\
-i(1-\Delta_{p_f}) & -iq^2(1-\Delta_{p_f}) & -i & 0
\end{pmatrix}\, ,
\end{align}
where
\begin{align}
    \Delta_{p_f}=\frac{1}{2}\left(\frac{\overline{\Delta}_{p}}{p^2-q^2}-\kappa_{f}\,\chi_{f}\right)\,,\quad \ \ \ \Delta_{q_f}=-\frac{1}{2}\left(\frac{\overline{\Delta}_{q}}{p^2-q^2}-\kappa_{f}\,\phi_{f}\right) \,.
\end{align}
The results for the metric $g$ are the same as in GR, so that we will mainly show the results for the metric $f$. For non-vanishing $\Lambda_{G}$, the Weyl invariants $\,^{(f)}\psi_2$ are the sum of the terms
\begin{align}
    \psi_{f_1}=-\frac{{\kappa_{f}}^2}{16\,\pi\,C^2}\frac{(m_{f}+n_{f})}{\left(p-q\right)^{3}}\, ,\quad  \ \ \ \psi_{f_2}=-\frac{{\kappa_{f}}^2}{16\,\pi\,C^2}\frac{(e_{f}^{2}-\mathrm{g}_{f}^{2})}{\left(p-q\right)^{3}\left(p+q\right)}.\,
\end{align}
The parameters form both metrics, $m_{G},n_{G},e_{G},\mathrm{g}_{G}$, are mapped to the gauge theories parameters $\widetilde{m}_{G},\widetilde{n}_{G},\widetilde{e}_{G},\widetilde{\mathrm{g}}_{G}$. Then, for the metric $g$ we obtain the same as in GR, and for the metric $f$ we get
\begin{align}
    \nonumber S_{(f_1)}& =-\frac{2}{3}\frac{16\,\pi}{{\kappa_{f}}^2}\frac{\left(\widetilde{m}_{f}+\widetilde{n}_{f}\right)^{2}}{C^2\,\left(m_{f}+n_{f}\right)\,\left(p-q\right)} \, , \quad
    &&f^{(f_1)}_{A B}= \left(-2\, \frac{\widetilde{m}_{f}+\widetilde{n}_{f}}{C^2\,\left(p-q\right)^{2}}\right)\,o_{(A}l_{B)} \, , \\
    S_{(f_2)} &=\frac{2}{3}\frac{16\,\pi\,C^2}{{\kappa_{f}}^2}\frac{\left(\widetilde{e}_{f}^{2}-\widetilde{\mathrm{g}}_{f}^{2}\right)^{2}}{\left(e_{f}^{2}-\mathrm{g}_{f}^{2}\right)\,\left(p-q\right)\left(p+q\right)} \, , 
    &&f^{(f_2)}_{A B}= \left(2\,\frac{\widetilde{e}_{f}^{2}-\widetilde{\mathrm{g}}_{f}^{2}}{\left(p-q\right)^{2}\left(p+q\right)}\right)\,o_{(A}l_{B)}.
\end{align}
The potential for $A^{(g_1)}_{\mu}$ is the same as in GR, with an analogous expression for $A^{(f_1)}_{\mu}$ with the corresponding parameters. For the metric $g$, the scalar and gauge field follow the same equations with the sources as in GR. For the metric $f$ we have that the sources are given by
\begin{align}
    \nonumber \,^{(0)}J^{\nu}_{(f_2)}&\equiv-\frac{(\widetilde{e}_f^{2}-\widetilde{\mathrm{g}}_f^{2})}{\left(p^{2}-q^{2}\right)^{3}}\left[(p^{2}+q^{2})\delta^{\nu}{}_{\tau}-2\delta^{\nu}{}_{\sigma}\right],\,  \\
    \nonumber \widetilde{\Lambda}_{(f_2)}^{\nu}&=0,\,\\
    \nonumber \,^{(0)}j_{(f_2)}&\equiv\frac{4}{3}\frac{16\,\pi}{{\kappa_{f}}^2}\frac{\left(\widetilde{e}_f^{2}-\widetilde{\mathrm{g}}_f^{2}\right)^{2}}{\left(e_f^{2}-\mathrm{g}_f^{2}\right)^{2}}\left[C^2\varepsilon\,\,^{(0)}\rho_{(f_2)}^{\text{grav}}+2\gamma_{p} \,\frac{(e_f^{2}-\mathrm{g}_f^{2})}{\left(q+p\right)^{3}\left(p-q\right)^{3}}\right],\,\\
    \lambda_{(f_2)}&\equiv
    -\frac{8}{9}\left(\frac{16\,\pi\,C^2}{{\kappa_{f}}^2}\right)^2\frac{\left(\widetilde{e}^{2}_f-\widetilde{\mathrm{g}}_f^{2}\right)^{2}}{\left(e_f^{2}-\mathrm{g}_f^{2}\right)^{2}}\,\,^{(0)}\rho_{(f_2)}^{\text{grav}}\,\left(\frac{p^{2}q^{2}}{p^{2}+q^{2}}\, \Lambda\right),\,
\end{align}
where $\,^{(0)}\rho_{(f_2)}^{\text{grav}}={R_{f}}^{\tau}{}_\tau$\,. The equations of motion for these fields are (\ref{typeD_eom_1_bg}).

\section{Conclusions}\label{Sec_conclusions}

The decomposition of the Weyl spinor of certain solutions with a generalized Kerr-Schild ansatz in bigravity was studied. For ``Type {\bf D}'' solutions, we treated Kerr-Newman-(A)dS and Pleba\'nski-Demia\'nski metric, and no massive modes are presented. For ``Type {\bf N}'' solutions, we studied pp-waves in vacuum and Siklos-(A)dS solution coupled electromagnetic and scalar matter and obtained massive modes contributing to the Weyl spinor, which leads to a massive prescription of the Weyl double copy. 

The equations of motion for type {\bf D} solutions are Maxwell equations coupled to an external source for the gauge theory and massless Klein-Gordon with a conformal curvature term and an external source. For type {\bf N} solutions, we obtained Proca equations coupled to a conformal term for the gauge theory and massive Klein-Gordon equations for the scalar theory.

The study of other solutions in massive bimetric gravity may be of interest, specially when considering solutions outside the Kerr-Schild family, or the consideration of additional fields in the theory.

\acknowledgments
It is a pleasure to thank E. Ayón-Beato and A. Luna for enlightening discussions. C. Ramos thanks SECIHTI for the scholarship No. 833288.


\end{document}